\documentclass[iop]{emulateapj} 

\usepackage{natbib}
\usepackage{graphicx}
\usepackage{amsmath}
\usepackage{color}
\usepackage{ulem}
\graphicspath{{./figs/}}

\newcommand{\Msun}{\rm{ M_{\odot}}}

\newcommand{\Mstar}{\rm M_{\rm{star}}}

\newcommand{\lya}{\rm {Ly{\alpha}}}

\newcommand{\Mh}{M_{\rm h}}
\newcommand{\Msunyr}{\rm M_{\odot}~ yr^{-1}}

\newcommand{\nh}{n_{\rm H}}

\newcommand{\sfr}{{\rm SFR}}
\newcommand{\rshalf}{R_{\rm star}^{\rm half}}

\begin{document}                          
%----------------------------------------------------------------------
%
% Title 
%
%----------------------------------------------------------------------

\title{Growth of First Galaxies:  Impacts of Star formation and Stellar Feedback}    

%----------------------------------------------------------------------
%
% Authors
%
%----------------------------------------------------------------------
\author
{
Hidenobu Yajima$^{\dagger}$\altaffilmark{1,2},
Kentaro Nagamine\altaffilmark{3,4},
Qirong Zhu\altaffilmark{5}, 
Sadegh Khochfar\altaffilmark{6},
Claudio Dalla Vecchia\altaffilmark{7, 8}
}

\affil{$^{1}$Frontier Research Institute for Interdisciplinary Sciences, Tohoku University, Sendai 980-8578, Japan}

\affil{$^{2}$Astronomical Institute, Tohoku University, Sendai 980-8578, Japan}

\affil{$^{3}$Theoretical Astrophysics, Department of Earth \& Space Science, Graduate School of Science, Osaka University, 1-1 Machikaneyama, Toyonaka, Osaka 560-0043, Japan}

\affil{$^{4}$Department of Physics and Astronomy, University of Nevada, Las Vegas, 4505 S. Maryland Pkwy, Las Vegas, NV 89154-4002, USA}

\affil{$^{5}$Department of Astronomy \& Astrophysics, The Pennsylvania State University, 
525 Davey Lab, University Park, PA 16802, USA}

\affil{$^{6}$Institute for Astronomy, University of Edinburgh, Royal Observatory, Edinburgh, EH9 3HJ, UK} 

\affil{$^{7}$Instituto de Astrof\'\i{}sica de Canarias, C/ V\'\i{}a L\'actea s/n, 38205 La Laguna, Tenerife, Spain}

\affil{$^{8}$Departamento de Astrof\'\i{}sica, Universidad de La Laguna, Av.~del Astrof\'\i{}sico Francisco S\'anchez s/n, 38206 La Laguna, Tenerife, Spain}

\email{$\dagger$ yajima@astr.tohoku.ac.jp} 

%----------------------------------------------------------------------
 
%\begin{document}

\label{firstpage}

%----------------------------------------------------------------------
%
% Abstract
%
%----------------------------------------------------------------------
\begin{abstract}
Recent observations have detected galaxies at high-redshift $z \sim 6 - 11$, 
and revealed the diversity of their physical properties, from normal star-forming galaxies to starburst galaxies. 
To understand the properties of these observed galaxies, 
it is crucial to understand the star formation (SF) history of high-redshift galaxies 
under the influence of stellar feedback. 
In this work, we present the results of cosmological hydrodynamic simulations with zoom-in initial conditions, 
and investigate the formation of the first galaxies and their evolution towards observable galaxies at $z \sim 6$. 
We focus on three different galaxies which end up in halos with masses 
$\Mh = 2.4 \times10^{10}~h^{-1}\; \Msun$ (Halo-10), $1.6\times10^{11}~h^{-1}\;\Msun$ (Halo-11) and $0.7 \times10^{12}~h^{-1}\Msun$ (Halo-12) at $z=6$.
 Our simulations also probe impacts of different sub-grid assumptions, i.e., SF efficiency and cosmic reionization, on SF histories in  the first galaxies.
We find that star formation occurs intermittently due to supernova (SN) feedback at $z \gtrsim 10$, 
and then it proceeds more smoothly as the halo mass grows at lower redshifts.
Galactic disks are destroyed due to SN feedback, while galaxies in simulations with no-feedback or lower SF efficiency models can sustain galactic disk for long periods $\gtrsim 10~ \rm Myr$.
The expulsion of gas at the galactic center also affects the inner dark matter density profile.
However, SN feedback does not seem to keep the shallow profile of dark matter for a long period. 
Our simulated galaxies in Halo-11 and Halo-12 reproduce the star formation rates (SFR) and stellar masses of observed Lyman-$\alpha$ emitters (LAEs) at $z \sim 7-8$  fairly well given observational uncertainties.
In addition, we investigate the effect of UV background radiation on star formation as an external feedback source, 
and find that earlier reionization extends the quenching time of star formation due to photo-ionization heating, 
but does not affect the stellar mass at $z=6$. 
\end{abstract}

%----------------------------------------------------------------------
%
% Keywords
%
%----------------------------------------------------------------------
%\begin{keywords}
\keywords{methods: numerical -- galaxies: ISM -- galaxies: evolution -- galaxies: formation -- galaxies: high-redshift}
%\end{keywords}

%----------------------------------------------------------------------
%
% Section 1: Introduction
%
%----------------------------------------------------------------------
\section{Introduction}
Understanding galaxy formation is one of the most important goals of today's astronomy and astrophysics. 
Recent development of observational facilities has allowed us to detect distant galaxies 
in the early Universe \citep[e.g.,][]{Ono12, Shibuya12, Bouwens10, Bouwens15, McLure13, Finkelstein13, Zitrin15, Oesch16}.
The most distant galaxy confirmed by spectroscopy has reached $z=11.1$ \citep{Oesch16}, and 
recent deep surveys are providing large samples of galaxies in the era of cosmic reionization at $z \gtrsim 6$
\citep[e.g.,][]{Ouchi09b, Bouwens09, Finkelstein11, Oesch13, Oesch14, Konno14, Bowler15}.   
In the 2020s, the next generation telescopes, e.g., James Webb Space Telescope (JWST), Thirty Meter Telescope (TMT) and European Extremely Large Telescope (E-ELT) are going to study star formation in the first galaxies at $z \gtrsim 10$ focusing on %by observing the associated radiation, e.g., 
UV continuum, Lyman-$\alpha$ line, and H$\alpha$ line. 
Therefore, it is crucial to make predictions of SF histories within the current theoretical framework before the next generation telescopes become online. 
The central question is how the first galaxies formed and evolved to the observed galaxies at $z \gtrsim 6$. 
In addition, the first galaxies are also considered to be important contributor of ionizing photons for cosmic reionization \citep{Yajima09, Yajima11, Yajima14c, Paardekooper13, Paardekooper15, Wise14, Kimm14, Robertson15,  Ma15}. 
Therefore, understanding star formation in first galaxies would also provide indispensable information 
for the cosmic reionization history. 

%%%-----------------------
\begin{table*}
\begin{center}
\begin{tabular}{ccccccccc}
\hline
Halo ID &  $\Mh \;[h^{-1}~\Msun]$ & $m_{\rm DM}\;[h^{-1}~\Msun]$ & $m_{\rm gas}\;[h^{-1}~\Msun]$ & $\epsilon_{\rm min}\:[h^{-1}\rm~pc]$   & SNe feedback 
& $A$ &$z_{\rm re}$\\
\hline
Halo-10      &$2.4 \times10^{10}$   & $6.6\times10^{4}$ & $1.2\times10^{4}$ & 200  &ON & $2.5\times10^{-3}$&10\\
%\hline
Halo-11   & $1.6\times10^{11}$ & $6.6\times10^{4}$  & $1.2 \times10^{4}$ & 200   &ON & $2.5\times10^{-3}$& 10\\
%\hline
Halo-11-lowSF  & $1.6\times10^{11}$ & $6.6\times10^{4}$  & $1.2 \times10^{4}$ & 200   &ON & $2.5\times10^{-4}$& 10\\
%\line
Halo-11-noSN      & $1.6\times10^{11}$   & $6.6\times10^{4}$ & $1.2\times10^{4}$ & 200  &OFF & $2.5\times10^{-3}$&10 \\
%\line
Halo-11-eUVB      & $1.6\times10^{11}$   & $6.6\times10^{4}$ & $1.2\times10^{4}$ & 200 &ON & $2.5\times10^{-3}$& 18 \\
%\hline
Halo-12   & $7.5 \times10^{11}$ %HY:  in the last 3 snapshots, Mh of Halo-12 fluctuates, and changes from 6.97, 8.65 to 6.94 x 10^11 Msun, the above 7.5 is the mean of last 3 snapshots. 
& $1.1\times10^{6}$  & $1.8\times10^{5}$ & 200  &ON & $2.5\times10^{-3}$& 10\\
\hline
\end{tabular}
\caption{
Parameters of zoom-in cosmological hydrodynamic simulations: 
(1) $\Mh$ is the halo mass at $z=6$. (2) $m_{\rm DM}$ is a dark matter particle mass. (3) $m_{\rm gas}$ is the initial mass of gas particles.
(4) $\epsilon_{\rm min}$ is the gravitational softening length in a comoving scale.   
(5) $A$ is the amplitude factor of the star formation model based on the Kennicutt-Schmidt law.
(6) $z_{\rm re}$ is the reionization redshift when the UVB is turned on.
}
\label{table:halo}
\end{center}
\end{table*}
%%%-----------------------

Recent cosmological simulations have been revealing the formation mechanism of first galaxies gradually 
\citep{Pawlik11, Wise12a, Wise12b, Johnson13, Jeon14, Kimm14, Hopkins14, Romano-Diaz14, Yajima15a, Yajima15c}.
Recently developed SN feedback models, which require high-spatial resolution of $\lesssim 10~\rm pc$, have shown that the star formation could be quenched easily by strong galactic outflows from shallow gravitational potential wells \citep{Kimm14, Hopkins14}.  Star formation is quenched for a while after the outflow is launched, however, the gas falls back again into the galaxy, and star formation resumes. 
As a result, simulations have shown that the star formation in first galaxies occurs intermittently. 
The quenching time-scale can be longer than the dynamical time of galaxies when most of the gas is evacuated.

Recent wide-field galaxy surveys allow us to study the typical halo mass of galaxies. 
\citet{Ouchi10} analyzed the clustering of LAEs at $z=6.6$ in their large-volume survey, 
and indicated that the halo mass of LAE hosts is $\Mh \sim 10^{11}~\Msun$. 
The simulated galaxies in previous studies were mostly in low-mass halos of $< 10^{11}~\Msun$ at $z \ge 6$ \citep[e.g.,][]{Wise12a, Pawlik13, Hopkins14, Ricotti16}.
Other populations of observed galaxies, such as Lyman break galaxies (LBGs) and submillimeter galaxies (SMGs), are likely to be hosted by more massive halos $\sim 10^{12}~\Msun$ \citep{McLure09}.
Therefore, in order to understand the evolution of the first galaxies from their formation towards observed galaxies at $z\sim 6-7$, more massive halos have to be studied.

In this work, we focus on the evolution of galaxies in dark matter halos 
that evolve to  $\Mh \approx 10^{10} - 10^{12}~h^{-1}\Msun$ at $z=6$.
We use the same simulation code used in the  First Billion Year (FiBY) project \citep[e.g.,][]{Johnson13, Paardekooper15}.
The FiBY project studied galaxy formation at $z \gtrsim 6$ using uniform calculation boxes of $4-32$ \,Mpc on a side with a maximum spatial resolution of 234 pc (comoving gravitational softening). 
Our simulations have similar resolution, but an important difference is using zoom-in initial conditions that allow to study
 more massive galaxies at $z=6$ than those in the FiBY project by choosing a larger box size initially before zooming in. 
Massive galaxies at higher redshift live in deeper gravitational potential wells where the feedback effects can be weaker. 
Using this sample, we investigate the evolution of the first galaxies from high redshift to $z \sim 6$, 
and examine if star formation in our simulated galaxies can reproduce the observed features. 
In addition, prior to the completion of cosmic reionization, 
the redshift when galaxies first begin to be affected by the UV background (UVB) sensitively depends on their location \citep[e.g.,][]{Hasegawa13, Iliev14}. 
In particular, the first galaxies near starburst galaxies would evolve under a UVB earlier. 
We also investigate the effects of different reionization redshifts on galaxy evolution.

Our paper is organized as follows.
We describe our simulations and initial conditions in Section~\ref{sec:model}. 
In Section~\ref{sec:result}, we present 
the most important properties of our simulated galaxies including their star formation histories and stellar distributions. 
We also investigate the effect of stellar feedback on the dark matter distributions and the dependence of halo mass
on the SF histories.
In addition, we show the effects of external UV radiation on star formation. 
Finally, we summarize our main conclusions in Section~\ref{sec:summary}.

%----------------------------------------------------------------------
%
% Section 2:  Model and Method
%
%----------------------------------------------------------------------
\section{Simulation Setup and Models}
\label{sec:model}

\subsection{Code and Zoom-in Method}

We use a modified version of the smoothed particle hydrodynamics (SPH) code {\sc gadget-3} \citep{Springel05e}
that was previously developed in the {\it Overwhelmingly Large Simulations} (OWLS) project \citep{Schaye10}.
This code was extended and modified to include the treatment of population III (Pop III) star formation, Lyman-Werner feedback, non-equilibrium primordial chemistry, and dust formation/destruction in the {\it First Billion Year} (FiBY) project (e.g., Khochfar \& Dalla Vecchia et al. in prep.).
The impacts of these new physical processes on galaxy formation 
have been studied in a series of papers of the FiBY project \citep[e.g.,][]{Johnson13, Paardekooper13, Agarwal15, Elliot15}. 
In this work, we do not consider the formation of Pop III stars and non-equilibrium primordial chemistry,
because our focus are more massive galaxies after the era of first mini-halos. 
We here use the star formation model based on the Kennicutt-Schmidt law, the supernova feedback and the equilibrium radiation cooling including metal lines as explained below. 

Our simulations follow the dynamics of dark matter and gas from $z=100$ to $z\sim 6$.
The initial conditions are created by the {\small MUSIC} code \citep{Hahn11}.
We first run a coarse N-body simulation with $128^3$ particles, and identify the dark matter halos using the friends-of-friends (FOF) grouping algorithm. 
We then choose three halos with  $\Mh = 2.4 \times 10^{10}\,h^{-1}\Msun$ (Halo-10), $1.6 \times10^{11}\,h^{-1}\Msun$ (Halo-11) and $7.5 \times10^{11}\,h^{-1}\Msun$ (Halo-12) at $z=6$ for subsequent zoom-in simulations.  
The entire simulation box size  is (20\,$h^{-1}$Mpc)$^3$ for Halo-10 and Halo-11,  
and (100\,$h^{-1}$Mpc)$^3$ for Halo-12 in a comoving units. 
For Halo-10 and Halo-11, the effective resolution is $2048^{3}$
and the zoom-in region is $\sim (1-2\,h^{-1}{\rm Mpc})^{3}$ in comoving scale.  
For Halo-12, the effective resolution is $4096^{3}$
and the zoom-in region is $(6.6\,h^{-1}{\rm Mpc})^{3}$.  
The relevant simulation parameters are summarized in Table~\ref{table:halo}.
In this work, we set the constant gravitational softening length of $200~h^{-1}~\rm pc$ in comoving scale
and the minimum smoothing length of $20~h^{-1}~\rm pc$. 
This corresponds to a resolution of $\lesssim 10 ~\rm pc$ physical at $z \sim 10$, which allows to resolve the internal gas structure for galaxies.

\begin{figure}
\begin{center}
\includegraphics[scale=0.55]{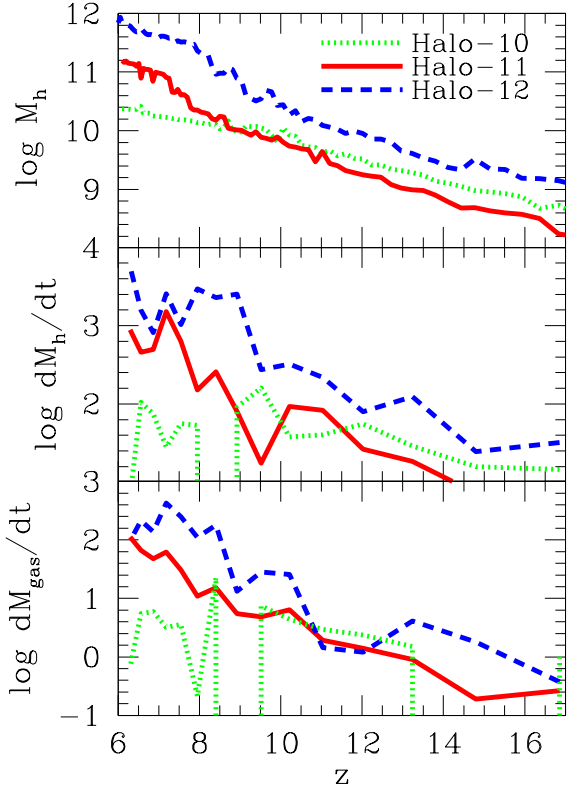}
\caption{
{\it Top panel}: Growth history of halo mass. 
Green dot, red solid and blue dash lines show the halo masses of the most massive halo as a function of redshift in Halo-10, Halo-11, and Halo-12 runs, respectively. 
{\it Middle panel}: Growth rate of halo mass in the unit of $\Msun~yr^{-1}$. 
{\it Bottom panel}: The rate of change of total gas mass within virial radius of the halo in the unit of $\Msun~yr^{-1}$.
The growth rates, $d\Mh/dt$ and $dM_{\rm gas}/dt$ are smoothed over the time bin $\Delta t=47~\rm Myr$.
}
\label{fig:hmass}
\end{center}
\end{figure}

Figure~\ref{fig:hmass} shows the redshift evolution of the halo mass, the growth rate of halo mass, and the rate of change of total gas mass. All halos grow as redshift decreases, although the growth rate depends on the environment.
Halo-10 is more massive than Halo-11 at $z \gtrsim 10$, however, Halo-11 rapidly grows and becomes more massive than Halo-10 during $z=6-9$.  Halo-12 is always more massive than the other halos. 
As shown in the middle panel of the figure, the halo growth rate of Halo-10 is higher than that of Halo-11 at $z \gtrsim 10$, but it stalls at lower redshift, whereas Halo-11 rapidly increases its growth rate at $z\lesssim 10$.

%%Fig.1 

\subsection{Star Formation}

Our basic SF prescription is based on the Kennicutt-Schmidt law of local galaxies, 
and the adopted formulation was developed by \citet{Schaye08}. 
This model estimates the SFR based on the local ISM pressure as follows:
\begin{equation}
\dot{m}_{*} = m_{\rm g}A 
\left( 1~\rm \Msun~pc^{-2} \right)^{-n} 
\left(  \frac{\gamma}{G} f_{\rm g} P \right)^{(n-1)/2},
\end{equation}
where $m_{\rm g}$ is the mass of the gas particle, $\gamma= 5/3$ is the ratio of specific heats,
$f_{\rm g}$ is the gas mass fraction in the self-gravitating galactic disk,  and $P$ is the total ISM pressure. 
The free parameters in this SF model are the amplitude $A$ and the power-law index $n$.
These parameters are related to the Kennicutt-Schmidt law, 
\begin{equation}
\dot{\Sigma}_{*} = A \left(\frac{\Sigma_{\rm gas} }{ 1 \Msun~{\rm pc^{-2}}} \right)^{n}. 
\end{equation}
Local normal star-forming galaxies follow $A_{\rm local}=1.5\times10^{-4}~\rm \Msun~yr^{-1}~kpc^{-2}$
and $n=1.4$ for the Saltpeter IMF. 
Note that, the amplitude should be changed by a factor $1/1.65$ in the case of the Chabrier IMF, i.e., $A_{\rm local,Chab}=2.5 \times10^{-4}~\rm \Msun~yr^{-1}~kpc^{-2}$. 
\citet{Schaye10} reproduced the observed cosmic star formation rate density (SFRD) using cosmological SPH simulations with this SF model and the parameters $A=2.5 \times 10^{-4}~\rm \Msun~yr^{-1}~kpc^{-2}$ and $n=1.4$\citep[see also,][]{Schaye15}.  
On the other hand, this amplitude can be $\sim 10$ times higher for merging starburst galaxies \citep{Genzel10}. 
In addition, \citet{Tacconi13} suggested that the amplitude increases with redshift. 

In this work, we focus on high-redshift galaxies at $z \ge 6$ which experience frequent merging processes. 
Therefore, assuming the Chabrier IMF, we adopt the amplitude factor $A$ which is 10 times higher than the local normal star-forming galaxies, i.e., $A_{\rm fiducial}=2.5 \times 10^{-3} ~\rm \Msun~yr^{-1}~kpc^{-2}$ as our fiducial value. 
The SF time-scale with this amplitude factor is $\sim 10^{8}~\rm yr$ which is close to the dynamical time of the halos at $z \sim 10$.
As shown in \citet{Schaye10},  the cosmic SFRD is not so sensitive to the value of $A$ due to the self-regulation by stellar feedback.
We also compare the SF histories of galaxies with the lower value of $A_{\rm local,Chab}$. 
In this work, we set $f_{\rm g}=1$ and the threshold density of $n_{\rm H}=10~\rm cm^{-3}$ above which star formation occurs, similarly to the FiBY project.
Above the threshold density, we use an effective equation of state with an effective adiabatic index $\gamma_{\rm eff}=4/3$
with the normalized pressure $P_{0}/k_{\rm B}=10^{2}~\rm cm^{-3}~K$ \citep[see][for more details]{Schaye08}.

\subsection{UV background radiation}

The cooling rate is estimated from the assumption of the equilibrium state (collisional or photoionization equilibrium) of each metal species. 
We allow the cooling down to $\sim 100~\rm K$ which is close to the CMB temperature at the redshifts we are focusing on.  
The metal-line cooling is considered for each metal species using a pre-calculated table by {\sc cloudy} v07.02 code \citep{Ferland00}.  
In this work, we do not follow the formation and dissociation of hydrogen molecules \citep{Johnson13}.
At $z \lesssim 10$, galaxies are being irradiated by the UVB, and it penetrates into the gas with $n_{\rm H} < 0.01 ~\rm cm^{-3}$ which is the threshold density of the self-shielding derived by \citet{Nagamine10a} and \citet{Yajima12h} based on the radiative transfer calculations of the UVB.  
We switch from collisional to photoionization equilibrium cooling table once the UVB ionizes the gas \citep[see][for details]{Johnson13}. 
We use the UVB of \citet{Haardt01} in our simulations.

%%Fig.2
\begin{figure*}
\begin{center}
\includegraphics[scale=0.85,angle=-90]{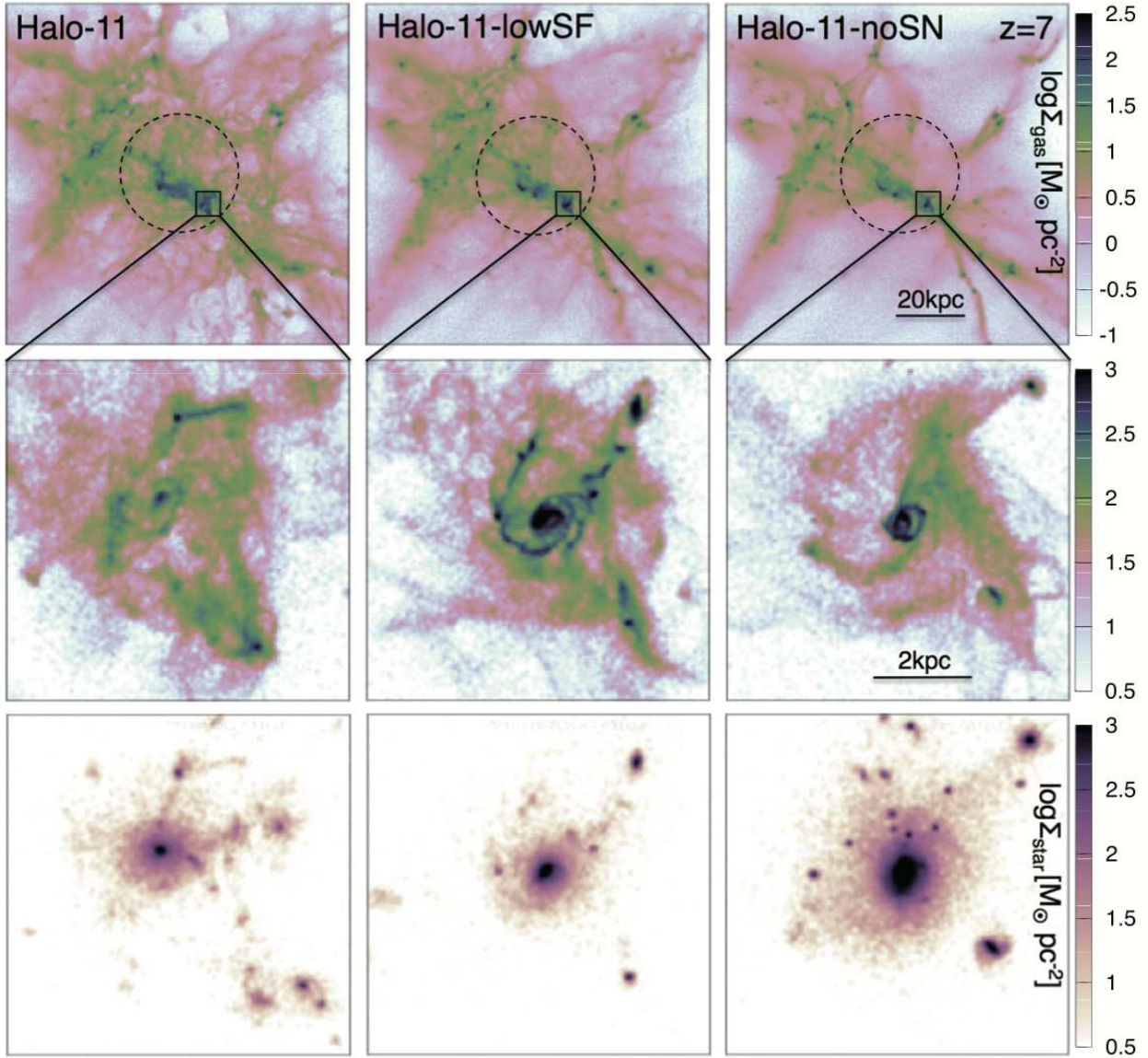}
\caption{
Column density of gas in Halo-11, Halo-11-lowSF, and Halo-11-noSN in units of [$\Msun\,pc^{-2}$] at $z=7$. 
See Table~\ref{table:halo} for the three variations of Halo-11.  
The top panels show the filamentary network of gas distribution on a scale of physical $100~\rm kpc$, 
and the dashed circles show the virial radius $\approx 18~\rm kpc$.
The middle panels show the zoom-in view of the most massive galaxy in the halo with a scale of physical $7~\rm kpc$.   
The lower panels represent the stellar surface density with the same logarithmic scale as the middle panels.
The galaxies have quite different morphologies depending on the treatment of SF and stellar feedback. 
}
\label{fig:map}
\end{center}
\end{figure*}

\subsection{Supernova Feedback}

In this work, we consider supernovae feedback via the injection of thermal energy into neighboring gas particles as described in \citet{DallaVecchia12}. 
The thermal energy is stochastically distributed to surrounding gas, and kept until the gas temperature increases to $10^{7.5}~\rm K$.
The feedback efficiency depends on the local physical properties, e.g., gas density, clumpiness, metallicity \citep[e.g.,][]{Cioffi88, Kim15}.
\citet{DallaVecchia12} compared the sound crossing time with the cooling time, and derived the following maximum gas density under which the thermal energy is efficiently converted to kinetic energy against radiative cooling loss:  
\begin{equation}
n_{\rm H} \sim 100~{\rm cm^{-3}} \left( \frac{T}{10^{7.5}~\rm K}\right) \left( \frac{m_{\rm g}}{10^{4}~\rm \Msun} \right)^{-1/2}.
\label{eq:nth}
\end{equation} 

In this work, we treat all star and gas particles as part of one galaxy system inside each halo 
that is identified by a FOF group finder, and do not distinguish sub-structures in each halo.  
Our current focus is the impact of stellar feedback on physical properties of galaxies, which 
sensitively depends on the SFR and gravitational potential of the dark matter halo. 
%\adm{\sout{Hence, we investigate the formation and evolution of \adb{the} first galaxies under the influence of star formation and its feedback together with \adb{the} dark matter halo mass assembly history using cosmological zoom-in hydrodynamic simulations.  }}
We adopt following cosmological parameters that are consistent with current cosmic microwave background observations: $\Omega_{\rm M}=0.3$, $\Omega_{\rm b}=0.045$, $\Omega_{\rm \Lambda}=0.7$, $n_{\rm s}=0.965$, and $h=0.7$ \citep{Komatsu11, Planck16}.  

%%%%%%%%%%%%%%%%%%%%%%%%%%%%%%%%%%%%%%%

%----------------------------------------------------------------------
%
% Section 3:  Results
%
%----------------------------------------------------------------------

\section{Results}
\label{sec:result}

\subsection{Star Formation History}
\label{sec:SFR}

The star formation history contains useful information on the details of galaxy formation and evolutionary processes. 
Recent observed galaxies at $z>6$ have been detected in the rest-frame UV, and their observed fluxes are directly linked to the SFR if dust extinction effects are negligible \citep[see however,][]{Yajima15c, Cullen17}. 

%In the next three subsections, 
Since the normalization factor $A$ is treated as a free parameter in our study,
we first examine the differences between the three runs of Halo-11, Halo-11-noSN and Halo-11-lowSF. 

Figure~\ref{fig:map} presents the maps of projected gas density for the Halo-11 with three different models at redshift $z=7$.
 In the case of Halo-11 in our fiducial model, the gas structure shows 
highly inhomogeneous and clumpy features. 
No extended galactic gas disk is seen within the central $\sim 2$ kpc of Halo-11. %in the middle left panel for Halo-11. 
On the other hand, disk-like structures are seen in Halo-11-noSN and Halo-11-lowSF models. 
This suggests the SF efficiency or the SN feedback have a large impact on the internal gas distribution of galaxies.  
The star formation is rather concentrated in the central nuclear region for the Halo-11-noSN, but the stellar distribution is more extended than the other runs due to larger amount of stars produced by $z=7$. 
%The SF efficiency or the SN feedback also 
As a result, both SF efficiency and SN feedback
affect the stellar distribution as shown in the lower panels of the figure. 
These effects will be discussed quantitatively in Section~\ref{sec:morphology}.

\begin{figure}
\begin{center}
\includegraphics[scale=0.45]{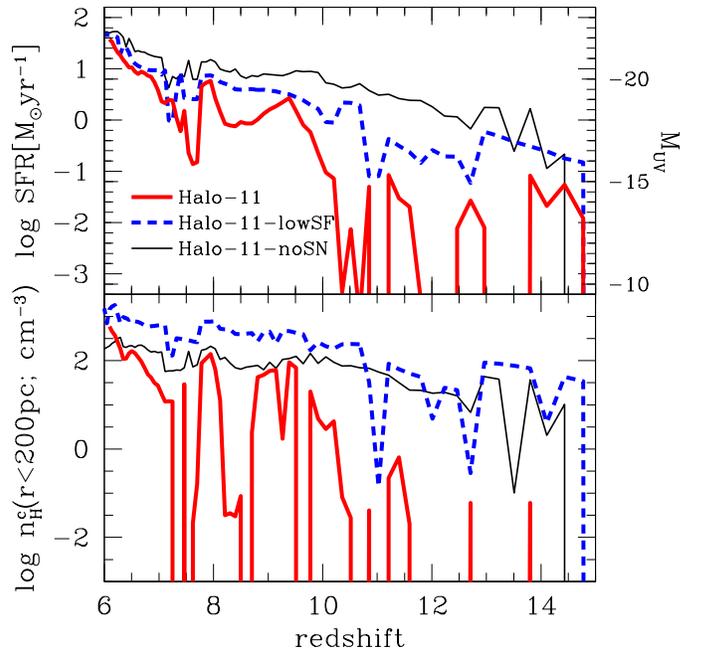}
\caption{
{\it Upper panel}: Star formation histories of simulated galaxies. 
Red solid, blue dash and black solid lines represent SFR of Halo-11, Halo-11-lowSF and Halo-11-noSN runs.
{\it Lower panel}: Volume-mean hydrogen number density within 200\,pc (physical) 
from the center of a galaxy as identified by the location of the maximum density in dark matter. 
}
\label{fig:sfr}
\end{center}
\end{figure}

Figure~\ref{fig:sfr} shows the SF histories of the three simulated galaxies. 
In the case without SN feedback (Halo-11-noSN), the SFR increases smoothly as the halo grows 
from $\sim 5~\Msunyr$ at $z \sim 10$ to $\sim 51~\Msunyr$ at $z\sim6$.
Using the relation $L_{\rm \nu, UV} = 0.7 \times 10^{28} \times ({\rm SFR / \Msunyr})~\rm erg \; s^{-1}$ \citep{Madau98, Kennicutt98}, 
we can convert the SFR to AB magnitude at UV band as follows: 
\begin{equation}
M_{\rm UV} = -18.0 - 2.5 {\rm log_{10}} \left( \frac{\rm SFR}{\Msunyr} \right).
\end{equation}
For example, based on this equation, the simulated galaxies reaches $-21.9, -22.3, -22.2$ mag at $z=6$ 
in Halo-11, Halo-11-noSN, Halo-11-lowSF. 
The detection limit of recent galaxy surveys at $z \lesssim 8$ has reached $\sim - 18$ mag \citep[e.g.,][]{Bouwens15}.
Therefore, if SN feedback is inefficient, galaxies hosted in halos with $\Mh \gtrsim 10^{11}~\Msun$ are likely to be detected easily as bright LBGs \citep[e.g.,][]{McLure13, Oesch14, Bouwens15}.  
\citet{Yajima15c} also show that massive galaxies formed in an overdense region are very bright. 
Note that, however, these authors have suggested that some fraction of UV radiation would be absorbed by interstellar dust \citep[see also,][]{Yajima14b, Cullen17}. 
In this case, galaxies will be bright at rest frame FIR wavelengths, %becomes bright at FIR band in rest frame, 
and can be observed by ALMA in the submillimeter wavelengths. 
In the case of Halo-11-lowSF, the star formation continuously occurs, but the SFR at $z \gtrsim 10$ is lower than that of Halo-11-noSN by a factor of a few to ten due to stellar feedback. 
The difference of SFR between Halo-11-lowSF and Halo-11-noSN becomes small as redshift decreases,
and is within a factor of $2$ at $z \lesssim 7$. 

%%Fig.3

On the other hand, the star formation of Halo-11 proceeds intermittently. 
At $z \gtrsim 10$, star formation is almost completely quenched due to stellar feedback
after experiencing a star formation event with SFR $\sim 0.1~\Msunyr$ in Halo-11. 
As the SFR decreases, SN feedback ceases to work.   
This allows the gas to fall down towards the galactic center, which leads to the next starburst. Repeating this cycle causes the oscillation of SFR.
The peak and the minimum of SFR is different by a factor of $\gtrsim 2$ dex at $z \gtrsim 10$. 
The mean SFR increases with decreasing redshift as the halo mass increases.  
Even with the high SF efficiency and SN feedback, the SFR becomes similar to Halo-11-noSN at $z \lesssim 7$. 
The SFR exceeds $1~\Msunyr$ at $z \lesssim 8$ and hence its UV flux can be higher 
than the detection limits of recent galaxy surveys for $z = 6 \sim 8$ \citep[e.g.,][]{Ouchi09b, Bouwens15}. 
At $z=6$, the SFR of Halo-11 reaches $\sim 37~\Msunyr$, whereas that of Halo-11-noSN is $\sim 51~\Msunyr$.
The Halo-11-lowSF run shows higher SFR than Halo-11 regardless of the lower SF efficiency. 
The SFR of Halo-11-lowSF is always between Halo-11 and Halo-11-noSN. 
This suggests that SFR is not simply proportional to the SF efficiency when the feedback is considered.
We find that the SFR of Halo-11-lowSF is continuous even at $z \gtrsim 10$ unlike Halo-11.
The low SF efficiency in the Halo-11-lowSF run prevents complete evacuation of gas from star-forming regions.

\subsection{Central Gas Density of First Galaxies}

The lower panel of Figure~\ref{fig:sfr} shows the time evolution of gas density within 200 pc (physical) 
from the center of the halo, which we take as the location of the maximum density in dark matter. 
The gas density decreases to even $n_{\rm H} < 10^{-3}~\rm cm^{-3}$ after the peak of SFR due to SN feedback. 
For example, at $z \sim 10$, the simulated galaxy has $SFR\sim 0.1-1~\Msunyr$ and the surrounding mean gas density is $n_{\rm H} \sim 10-100~\rm cm^{-3}$. 

Let us examine in the following whether SN feedback can disrupt these dense gaseous clumps in high-$z$ galaxies. 
Given that the life-time of massive stars (which causes the core-collapse SNe) is less than $ 10~\rm Myr$, 
the released SN energy results in $E \sim 10^{55}~\rm erg$ for $SFR=0.1~\Msunyr$.
Here we consider the energy of each supernova as $10^{51}~\rm erg$, and estimate the number of supernova per one solar mass of stars formed as $\sim 10^{-2} \times M_{\rm star} / \Msun$, where $M_{\rm star} = SFR \times t_{\rm life}$. 
The Jeans mass is $\sim 10^{7}~\Msun$ for the condition of $n_{\rm H} \sim 100~\rm cm^{-3}$ and $T \sim 10^{4}~\rm K$. 
The gravitational binding energy of the spherical gas cloud induced by the Jeans instability is $E \sim 2GM_{\rm Jeans}^{2}/ L_{\rm Jeans} \sim 10^{53}~\rm erg$, which is lower than the released SN energy for $SFR=0.1~\Msunyr$.
Therefore, if the injected thermal energy is efficiently converted into the kinetic energy, 
the star-forming clouds of $\sim 10^{7}~\Msun$ can be destroyed. 
Halo-11-noSN and Halo-11-lowSF always retain the high-density gas near the center with $n_{\rm H} \ge 100~\rm cm^{-3}$ at $z < 10$. 
This suggests that the disruption of star-forming clouds depends on the SF efficiency. 

The slow SF model of Halo-11-lowSF allows the gas to concentrate at the galactic center,  resulting in the formation of high-density gas cloud. 
In the high-density region, the gravitational binding energy is high due to high-density gas clumps populating the small volume.
In addition, the SN feedback efficiency becomes weaker due to radiative cooling loss, and the conversion efficiency from thermal to kinetic energy depends on the local gas density as discussed in Section~\ref{sec:model}. 
Thus, Halo-11-lowSF cannot evacuate the gas near the galactic center unlike Halo-11. 
Note that the gas density of Halo-11-lowSF exceeds that of Halo-11-noSN at $z \lesssim 10$. 
As will be shown in Section~\ref{sec:shmr}, a large fraction of gas in Halo-11-noSN  is efficiently converted into stars even at $z \gtrsim 6$, resulting in a lower gas mass fraction at the galactic center compared to the Halo-11-lowSF.

%%KN: this following subsection suddenly starts to discuss DM profile and BH.   Maybe move to later sections, e.g. Discussion section?
%%HY: I have moved this sentence to Summary section.

%The rapid change in gravitational potential can affect the dark matter density profile as we discuss in Section~\ref{sec:dmprof}. 
%In addition, the gas density at the galactic center also affects the co-evolution of central massive black hole (BH) and the stellar bulge. 
%\citet{Dubois15} recently demonstrated that SN feedback can eject the gas from galaxies, and significantly suppress the growth of the central BH. 
%Our simulations indicate low gas densities near the galactic center, which suggests slower growth of BHs for those galaxies that have been affected strongly by the SN feedback. 

\subsection{Gas Content of First Galaxies}

\begin{figure}
\begin{center}
\includegraphics[scale=0.42]{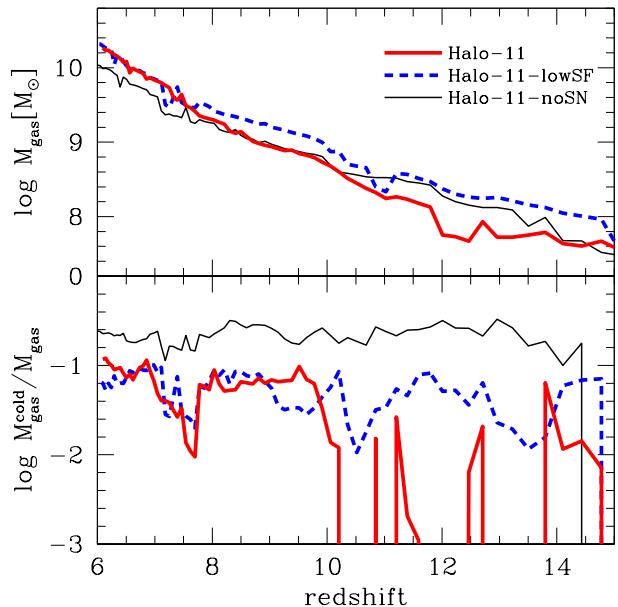}
\caption{
{\it Upper panel}: Redshift evolution of total gas mass in three different runs of Halo-11. 
{\it Lower panel}: The mass fraction of cold, high-density gas with $n_{\rm H} \ge 10~\rm cm^{-3}$ and $T\le10^{4}~\rm K$ to the total.
}
\label{fig:coldgas}
\end{center}
\end{figure}

Figure~\ref{fig:coldgas} presents the redshift evolution of total gas mass and cold gas mass fraction. 
Here we define the cold gas as the one with temperature less than $10^{4}~\rm K$ and hydrogen number density higher than $10~\rm cm^{-3}$. %%KN:   did you also impose density cut according to the caption of Fig. 4?   %%HY: added.
Total gas mass monotonically increases as the halo grows. 
The gas mass of Halo-11 is lower than the other runs  at $z \gtrsim 10$,  
and the difference becomes larger at higher redshift, although they are always within a factor of few. 
This suggests that the gas mass evacuated from a halo due to the feedback is insignificant even in the case of Halo-11.  
On the other hand, as shown in the lower panel of the figure, the cold gas mass shows a large difference among different runs. 
At high-redshift, even if accreted gas experiences the virial shock, it can quickly cool down via efficient atomic line cooling
and settle into the central galactic disk. Thus, in the case of Halo-11-noSN, more than $10~\%$ of gas always stays in the cold high-density phase. 
Once stellar feedback becomes effective, the gas can be heated and evacuated from star-forming regions. 
Halo-11-lowSF can only retain $1-10$\% of the total in cold phase. 
In the case of Halo-11, at $z \gtrsim 10$, most of gas is evacuated from star-forming regions and becomes low-density hot phase
after the short, active star formation. 
Then, even the efficient star formation with feedback in Halo-11 cannot erase all cold, high-density gas
at $z \lesssim 10$ when the halo mass exceeds $\sim 0.7 \times 10^{10}~h^{-1}~\Msun$.
In this phase, although most of the gas is evacuated from the galactic central region as shown in Figure~\ref{fig:sfr}, 
cold sub-clumps that are distributed far from the galactic center keep star formation going.  
This makes the continuous SF history of Halo-11 at $z \lesssim 10$.

%As star formation proceeds, type-II SNe produce metals and dust. 
%Figure~\ref{fig:metal} represents the redshift evolution of total stellar mass and metallicity of Halo-11. 
%As the mean SFR increases, the stellar mass rapidly increases, 
%and reaches $\Mstar = 2.5 \times 10^{9}~\Msun$ at $z=6$ for Halo-11. 
%The majority of  stars form at $z \lesssim 8$.  
%The metallicity reaches $\sim 1\times10^{-2}~\Zsun$ at $z\sim10$ and $\sim 4 \times 10^{-2}~\Zsun$ at $z \sim 6$ for Halo-11. 
%Note that, unlike the stellar mass, the metallicity oscillates as a function of redshift. 
%This is because the metal-enriched gas is blown out from galaxies, and then primordial intergalactic medium is accreted. 
%Through the cycle of gas inflow and outflow, the metallicity changes rapidly with redshift. 
%As the halo grows, metal enriched gas gradually accumulates in the galaxy, and the mean metallicity increases with decreasing redshift. 

%%Fig. 4

%\begin{figure}
%\begin{center}
%\includegraphics[scale=0.43]{metal.eps}
%\caption{
%{\it Upper panel}: Redshift evolution of stellar mass. 
%Red solid and black dash lines represent Halo-11 and Halo-11-noSN, respectively.
%{\it Lower panel}:  Gas-phase metallicity as a function of redshift in the same halos. 
%}
%\label{fig:metal}
%\end{center}
%\end{figure}

%%%%%%%%%%%%%%%%%%%%%%%%%%%%%%%%%%

\subsection{Halo Mass Dependence of SFR and Stellar Mass}
\label{sec:halomass}

%Fig.5
\begin{figure}
\begin{center}
\includegraphics[scale=0.45]{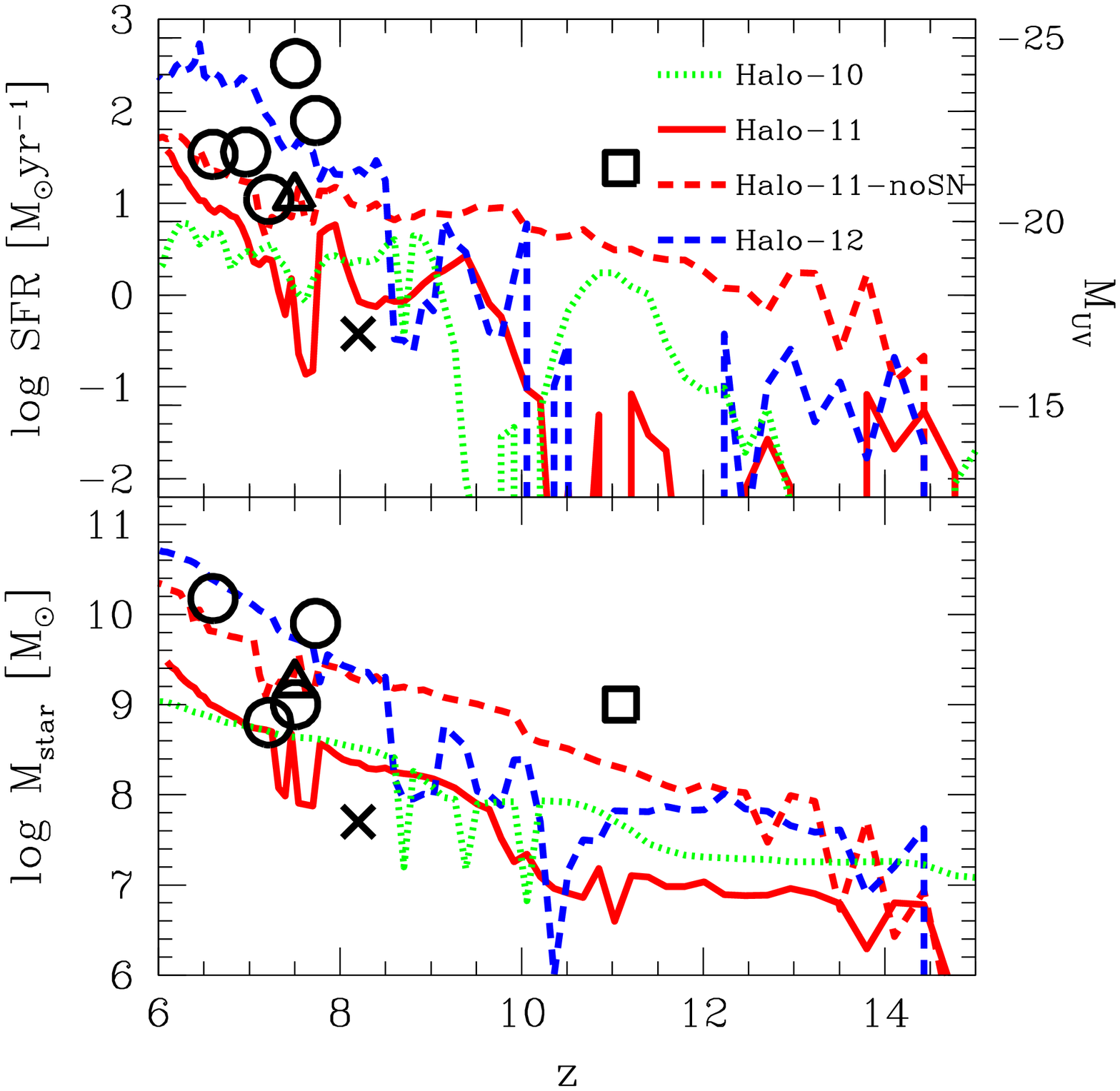}
\caption{
Star formation rates and stellar masses in different halos compared with observational data. 
Open circles represent $\lya$ emitting galaxies at $z=6.6$ \citep{Ouchi09b}, 
$z=6.96$ \citep{Iye06}, $z=7.213$ \citep{Ono12}, $7.3$ \citep{Shibuya12},
$7.51$ \citep{Finkelstein13}, $7.73$ \citep{Oesch15}, and $z=8.68$ \citep{Zitrin15}.
The open square shows the Lyman-break galaxy at $z=11.09$ \citep{Oesch16}. 
The open triangle represents the sub-millimeter galaxy at $z=7.5$ \citep{Watson15}.  
The cross indicates the host galaxy of GRB at $z=8.2$ \citep{Tanvir12}.
}
\label{fig:sfrstm}
\end{center}
\end{figure}

Next we study the evolution of first galaxies in halos with different masses.
In addition to the three models of Halo-11, we also follow two other halos: 
one with $\Mh=2.4 \times 10^{10}~\Msun$ at $z=6$ (Halo-10),
and another with $\Mh=0.7 \times 10^{12}~\Msun$ (Halo-12). 
%\adm{\sout{For Halo-12, we simulate a cosmological box of $L=100~\rm Mpc$,  and construct a zoom-in initial condition which yields a massive halo. }}

Figure~\ref{fig:sfrstm} shows the SFR and stellar mass of different halos
together with recent observational data of spectroscopically confirmed galaxies. 
We find that a more massive halo (Halo-12) shows a rapid variation in SFR, similarly to Halo-11 at $z \gtrsim 10$. 
After the star formation at $z = 12.2$, the SN feedback evacuates the gas and quenches the star formation for a while until $z=10.5$.
As halos become more massive towards lower redshift, the deeper gravitational potential well can hold the gas against SN feedback, which allows for more continuous SF at $z \lesssim 10$. 
We find that Halo-11 and Halo-12 show similar SFRs as the observed LAEs at $z \sim 7$ and the SMG at $z=7.5$ discovered by ALMA \citep{Watson15}. 
As shown in Figure~\ref{fig:hmass}, the halo mass of Halo-11 and Halo-12 is in the range of $\sim  9 \times 10^{10} - 4 \times 10^{11}~\Msun$ at $z \sim 7$. 
These halo masses are consistent with the results of semi-analytic model of LAEs augmented by radiation transfer calculation \citep{Yajima17}, and the clustering analysis of LAEs at $z \sim 7$, which indicated a typical host halo mass of $\sim 10^{11}~\Msun$ \citep{Ouchi10}. 

Halo-10 has a somewhat higher halo mass than Halo-11 at $z \gtrsim 10$, 
therefore Halo-10 has a similar or higher SFR and a higher $\Mstar$ than Halo-11 at those redshifts. 
Then, as the redshift decreases, the mass growth of Halo-10 becomes slower than Halo-11, resulting in lower SFRs at $z \lesssim 7$. 
As a result, the SFR of Halo-10 cannot reach the observed SFR of LAEs. 
Halo-12 shows high SFRs of $\gtrsim 10~\Msunyr$ at $z \lesssim 8$, which is similar to the bright LAEs at  $7.73$ \citep{Oesch15}. 

It is interesting that none of our simulated galaxies reproduce the observed SFR of GN-z11 ($\sfr = 24~\Msunyr$) which is the most distant galaxy observed to date \citep{Oesch15}.
Halo-12 shows only ${\rm SFR} \lesssim 1~\Msunyr$ at $z \gtrsim 10$ even in its starburst phase. 
In order to reproduce the high SFR, other conditions may be required, 
e.g., more massive halos or different halo growth histories. 

The lower panel of Figure~\ref{fig:sfrstm} shows the evolution of stellar mass. 
At first, Halo-10 has a higher $\Mstar$ than Halo-11 at $z > 10$. 
Then, as the halo growth stalls at $z<10$ (see Fig.~\ref{fig:hmass}), Halo-10 does not increase its $\Mstar$ very much, 
only reaching $\Mstar \sim 1 \times 10^{9}~\Msun$ at $z \sim 6$. 
On the other hand, Halo-11 increases its $\Mstar$ smoothly as redshift decreases, 
and reach $\Mstar \sim 3\times10^{9}~\Msun$ at $z \sim 7$, which matches the observed LAEs. 
Higher SFRs of Halo-12 lead to higher $\Mstar$ than in Halo-11, achieving 
$\Mstar \gtrsim 3 \times 10^{9}~\Msun$ at $z \lesssim 8$, which agrees with the bright LAE at $z=7.73$. 
Sometimes the stellar mass drops suddenly, but this is due to galaxy merger events that spread the stellar components
beyond the virial radius.  In this work, we only take stars within the virial radius into account, and it sometimes lead to decreasing stellar masses. 
Note that, if angular resolution of future telescopes will be able to resolve substructures with a scale of virial radius $\sim 10~\rm kpc$,  galaxies in such a merger event can be identified separately. 
The detailed modeling of observational properties of first galaxies will be investigated with radiation transfer calculations in our future work.
Again, $\Mstar$ of Halo-12 is lower than GN-z11 by an order of magnitude at $z\sim 11$.  
The observed high $\Mstar$ indicates that GN-z11 kept high SFRs of $\gtrsim 1-10~\Msunyr$ over $t \gtrsim 10^{8}~\rm yr$,  and it is unlikely to be an instantaneously bright galaxy due to major merger if the observed stellar mass is correct. 

%%%%%%%%%%%%%%%%%%%%%%%%%%%%%%%

\subsection{Evolution of Stellar-to-Halo Mass Ratio (SHMR)}
\label{sec:shmr}

\begin{figure}
\begin{center}
\includegraphics[scale=0.4]{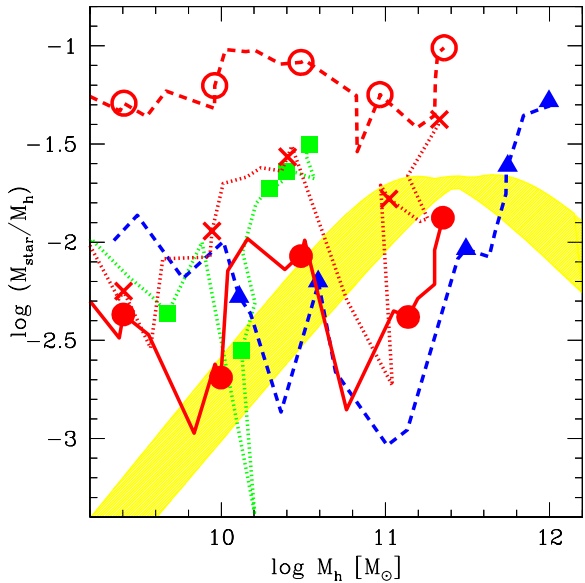}
\caption{
Redshift evolution of SHMR vs. halo mass is shown. 
Each symbol indicates different epochs of $z= 12, 10, 8, 7$ and $6$ from left to right for Halo-11 (red filled circle),  Halo-11-noSN (open circles), Halo-11-lowSF (red crosses), Halo-10 (green squares),
and Halo-12 (blue triangles). 
Yellow shaded region shows the abundance matching result by \citet{Behroozi13} at $z = 6-8$. 
}
\label{fig:mhmst}
\end{center}
\end{figure}

We now discuss the relation between stellar mass and halo mass. 
\citet{Behroozi13} derive a relation between halo and stellar mass based on the abundance matching analysis, 
and show that the stellar mass increases with halo mass until a specific mass in order to match the faint-end of the galaxy stellar mass function. 
Figure~\ref{fig:mhmst} shows the evolution of SHMR as a function of halo mass for different halos. 
The SHMR of Halo-11-noSN is higher than that of \citet[][yellow band]{Behroozi13} by one to two orders of magnitudes.
This means that Halo-11-noSN converts gas into stars too efficient, 
which resembles the canonical over-cooling problem seen in the hydrodynamic simulations \citep[e.g.,][]{Scannapieco12}. 
On the other hand, SHMR of Halo-10, Halo-11, and Halo-12 roughly matches that of \citet{Behroozi13} at later times. 
Therefore we confirm that stellar feedback is the key for regulating star formation and reproducing stellar masses.   
The SHMR of Halo-12 at $z=6$ is much higher than Behroozi's result, but the discrepancy can be understood by following possible reasons:
(1) the small sample of simulations: the SHMR derived from the abundance matching method is a statistical mean value. Our small sample might not follow the statistical mean perfectly.
(2) the limited sample of observed galaxies at $z \ge 6$: Due to the difficulty of estimating stellar mass of galaxies at $z\ge 6$, the estimation of \citet{Behroozi13} suffers from large uncertainties at $z \gtrsim 6$. 
(3) missing other feedback processes: our simulations consider only SN feedback and photo-ionization heating by UVB, and other processes, such as radiation pressure on dust and AGN feedback, may suppress SF in massive halos. 

%Fig.6

\subsection{Evolution on the Main Sequence of Star Formation}

\begin{figure}
\begin{center}
\includegraphics[scale=0.4]{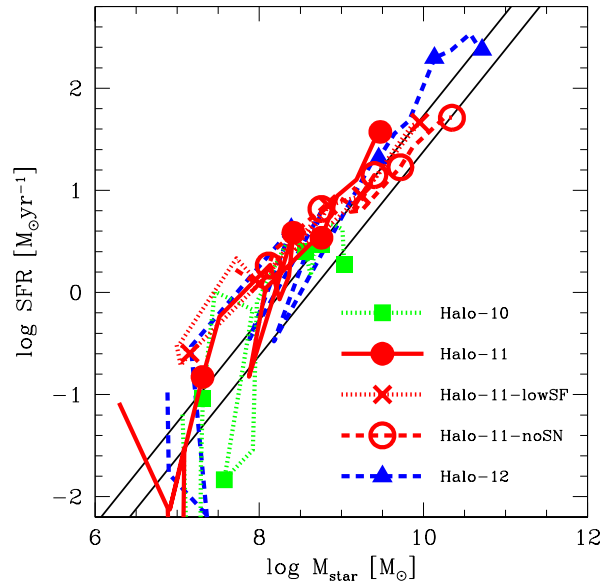}
\caption{
Main sequence of star formation ($M_{\rm star}$ vs. SFR). 
The symbols and line types are the same as in Fig.~\ref{fig:mhmst}.
Upper and lower black solid lines are the observational results at $z=3.8-5.0$ \citep{Marmol16} 
and $z=2.0$ \citep{Daddi07}. 
}
\label{fig:mainseq}
\end{center}
\end{figure}

The relation between SFR and $\Mstar$ is frequently discussed in the context of SF efficiency, and is often called the main sequence of star formation \citep{Daddi07}. 
Figure~\ref{fig:mainseq} shows the relation between SFR and $\Mstar$ for our simulated halos. 
We made the smoothed curves by taking the averaged SFR within a time bin of $\Delta t = 47~\rm Myr$. 
The solid lines are the observed main sequence at $z=3.8-5.0$ \citep[][upper black line]{Marmol16}  
and $z=2.0$ \citep[][lower black line]{Daddi07}.
Our simulated galaxies at $z \ge 10$ show higher SFR than the observed main sequence at $z=3.8-5.0$ by a factor of a few, 
then gradually approaches the observed lines towards $z \sim 6$. 
This indicates that the star formation proceeds efficiently at $z > 6$ in simulated galaxies due to higher gas fraction and higher gas density. 
Note that even Halo-11-noSN shows a similar curve to other runs with feedback. 
SN feedback suppresses the SFR temporarily (as partially seen in Halo-10), but it does not affect the mean SFR on a time scale $\Delta t = 47~\rm Myr$ for Halo-11 and Halo-12. 
Hence we suggest that the SF main sequence is not so sensitive to stellar feedback. 

\begin{figure}
\begin{center}
\includegraphics[scale=0.42]{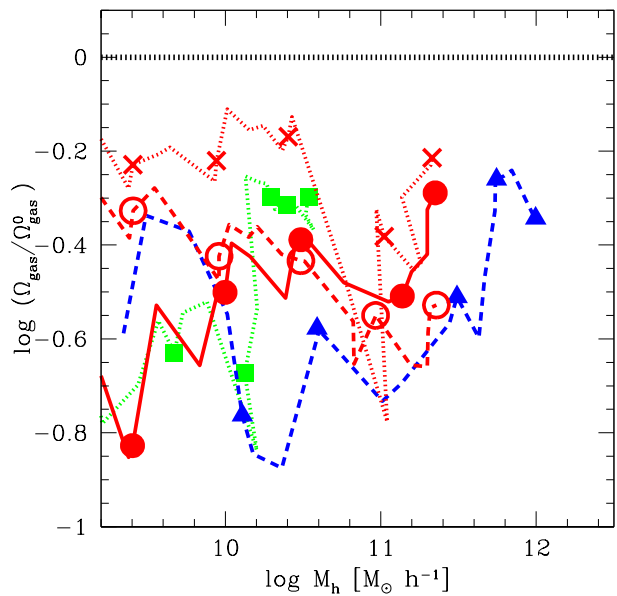}
\caption{
Total gas mass fraction as a function of halo mass, normalized by the cosmic mean value  ($\Omega_{\rm b}/\Omega_{\rm m}$). 
The symbols and line types are the same as in Fig.~\ref{fig:mhmst}.
Horizontal black dotted line indicates the cosmic mean value. 
}
\label{fig:mhmgas}
\end{center}
\end{figure}

As star formation proceeds, the gas mass fraction decreases due to outflow and conversion into stars. 
Figure~\ref{fig:mhmgas} shows the gas mass fraction to halo mass normalized by the cosmic mean value 
$\Omega_{\rm gas}^{\rm 0} \equiv \Omega_{\rm b}/\Omega_{\rm M} \approx 0.16$.  
We find that the gas mass fraction of Halo-10 and Halo-11 significantly decrease due to SN feedback
when their halo mass is lower than $\sim 10^{9} ~\Msun$. 
The minimum reaches around 20 per cent of the cosmic mean.  
Then, as halos grow and gravitational potential becomes deeper, even the gas kicked by SNe is trapped, 
i.e., outward velocity is less than the escape velocity of the halo. 
As a result, the gas mass fraction gradually increases and stays at $\sim 40$\% of the cosmic mean when $\Mh \gtrsim 10^{10}~\Msun$.
On the other hand, the gas mass fraction of Halo-12 is not so low even when $\Mh < 10^{9}~\Msun$.
The redshift when the halo mass reach $10^{9}~\Msun$ is $\sim 18$ for Halo-12, 
while $z\sim 14$ for Halo-11. Hence, the galaxy is more compact and can efficiently trap gas against SN feedback. 
At $z \lesssim 12$, the gas mass of Halo-12 is significantly reduced due to the feedback, 
and reaches $\sim 20$\% of the cosmic mean. 
Then, as redshift decreases, the gas mass fraction increases gradually, and becomes similar to that of Halo-11 when $\Mh \gtrsim 10^{11}~\Msun$.
On the other hand, the gas mass fraction of Halo-11-noSN monotonically decreases due to too efficient star formation, 
resulting in $\rm \Omega_{\rm gas}/\Omega_{\rm gas}^{0} = 0.3$ at $z\sim 6$.

%Fig.7

%Fig.8

%%%%%%%%%%%%%%%%%%%%%%%%%%%%%%%%%%

\subsection{Effect of SN feedback on galactic morphology}
\label{sec:morphology} 

The global structure of gas in a galaxy can change dramatically after the starburst and subsequent SN feedback.   
The standard picture of galaxy formation asserts that the warm gas with $T=10^4-10^5$\,K cools by radiative cooling, falls into the center of halo, and forms a rotationally-supported galactic disk \citep[e.g.,][]{Mo98}.
\citet{Pawlik13} argued that the first galaxies can maintain galactic disks against stellar radiation feedback. 
However, since the first galaxies are hosted in low-mass halos, SN feedback can significantly affect the global gas structure \citep[e.g.,][]{Kimm14}. 
The halo spin parameter of Halo-11 is $\lambda \sim 0.1$ at $z\sim6$, and if there is no feedback effects, it is likely to have an extended galactic disk with $R_{\rm disk} \sim \lambda R_{\rm vir}$, where $\lambda$ is the halo spin parameter in the range of $\sim 0.01 - 0.1$ \citep{Mo98}. 

\begin{figure}
\begin{center}
\includegraphics[scale=0.42]{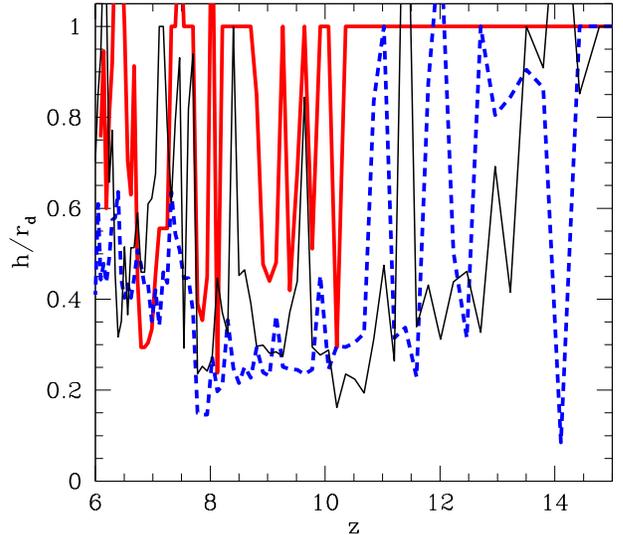}
\caption{
Thickness of galactic disk $h$ as a function of redshift.
The $z$-axis (normal direction to disk) is determined by aligning the angular momentum vector of all gas particles within physical 200\,pc from galactic center. 
The disk radius $r_{\rm d}$ is defined such that the mean density at the mid-plane within $r_{\rm d}$ becomes $n_{\rm H}=10~\rm cm^{-3}$. 
The thickness $h$ is the height where the mean density of the cylinder exceeds the same threshold density.
}
\label{fig:disk}
\end{center}
\end{figure}

Figure~\ref{fig:disk} shows the thickness of galactic disks, $h/r$.
Since the current numerical resolution is somewhat poor to fit the galactic disk with an exponential profile, 
here we roughly measure the thickness of galactic disk as follows. 
The $z$-axis (normal direction to the disk) is determined by aligning the angular momentum vector 
of all gas particles within physical 200 pc from the galactic center. 
The disk radius $r_{\rm d}$ is defined such that the mid-plane mean density becomes $n_{\rm H}=10~\rm cm^{-3}$ inside $r_{\rm d}$. 
The thickness $h$ is the height when the mean density of the cylinder with a radius of $0.2\, r_{\rm d}$ exceeds the same threshold density.
When the gas is evacuated from the galactic center as shown in Figure~\ref{fig:sfr}, $h/r_{\rm d}$ is set to unity for plotting purposes. 
Note that $h/r_{\rm d}$ can exceed unity when high-density gas distributes along the face-on viewing angle to the disk.
Due to gas expulsion, Halo-11 does not have a disk at $z \gtrsim 10$. 
Then, at $z \lesssim 10$, a galactic disk forms when gas fills in around the galactic center
as shown in Figure~\ref{fig:sfr}.  However, Halo-11 cannot maintain this disk for a long period due to SN feedback. 
On the other hand, Halo-11-noSN and Halo-11-lowSF show disky shapes with $h/r_{\rm d} \lesssim 0.5$ for longer periods. 
In the case of Halo-11-lowSF, a disk forms at $z \gtrsim 10$, but it is destroyed due to SN feedback. 
Then, as the halo grows, the deeper gravitational potential well allows galaxies to support their galactic gas disks against feedback. 
Thus, the $h/r_{\rm d}$ of Halo-11-lowSF always shows less than $\sim 0.6$ at $z \lesssim 10$.
Halo-11-noSN also has low values of $h/r_{\rm d}$ owing to no feedback. 
Unlike Halo-11-lowSF, the gas clumps can be efficiently converted into stars before they fall into the galactic center, 
resulting in higher values of $h/r_{\rm d}$ than for Halo-11-lowSF.

%Fig.9

\begin{figure}
\begin{center}
\includegraphics[scale=0.42]{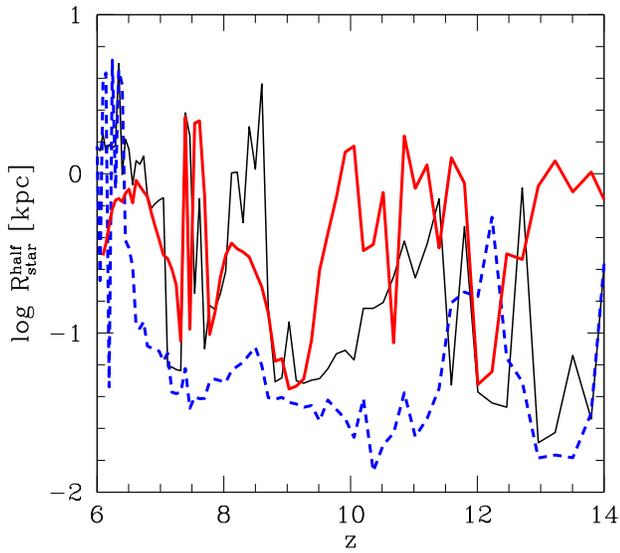}
\caption{
Half-mass radius of stars $R^{\rm half}_{\rm star}$ in physical units. 
Each line corresponds to Halo-11 (red solid), Halo-11-lowSF (blue dashed), and Halo-11-noSN (black solid) runs, respectively. 
}
\label{fig:msthalf}
\end{center}
\end{figure}

In addition, the evacuation of gas near the galactic center can lead to different stellar distribution between simulations with and without SN feedback. 
Figure~\ref{fig:msthalf} shows the half-mass radius of stars ($\rshalf$) as a function of redshift. 
We estimate the stellar mass within the half of virial radius $R_{\rm vir}(\Mh, z)$, and
measure $\rshalf$ where the stellar mass becomes half of $M_{\rm star} (<0.5 R_{\rm vir})$.
We first need to define the center of galaxies in order to compute $\rshalf$. 
When galaxies are in major merger processes, galaxies in subhalos can distribute far from the center of mass of a halo. 
Considering the large separations of substructures,
we define the galactic center as the position of highest density in dark matter substructure
and focus on a substructure with the scale of $0.5 R_{\rm vir}$.
One might naively expect that $\rshalf$ would be small if there is no SN feedback. 
However, even Halo-11-noSN shows large values of $\rshalf$ in spite of no feedback.
Some peaks of $\rshalf$ are caused by sub-components of star clusters distributing far from the center of galaxy. 
In the case of Halo-11, $\rshalf$ takes large values $\gtrsim 1~\rm kpc$, whereas that of Halo-11-noSN steeply decreases after each peak. 
As shown in Figure~\ref{fig:sfr}, the SN feedback evacuates gas from star-forming regions near the galactic center, 
and suppresses the concentration of stellar distribution. 
The slow star formation of Halo-11-lowSF allows continuous SF at the galactic center. 
In the case of Halo-11-lowSF, star formation in low-density regions far from galactic centers is inefficient, 
resulting in the lower $R_{\rm half}$ with $\lesssim 0.1~\rm kpc$ which corresponds to $\lesssim 1~\%$ of virial radius. 
These results indicate that the low SF efficiency is more important in making the compact stellar bulge rather than non-existence of SN feedback. 
Thus, we suggest that SF efficiency and SN feedback affect the compactness of stellar distribution, 
leading to different galaxy growth rates of galactic bulges. 
The angular resolution of MIRI imaging of JWST will be able to resolve the stellar distribution with the scale of $\sim 1~\rm kpc$.
Therefore these sub-grid models of feedback and SF efficiency can be investigated via the comparisons with future observations.

%Fig.10

\subsection{Phase Diagram}

\begin{figure}
\begin{center}
\includegraphics[scale=0.45]{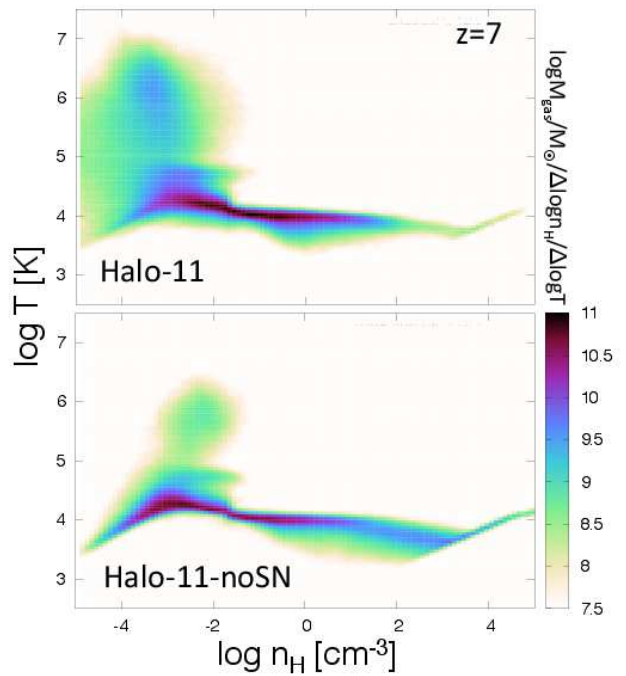}
\caption{
Phase diagram of gas in Halo-11 and Halo-11-noSN at $z=7$.
}
\label{fig:phase}
\end{center}
\end{figure}

Figure~\ref{fig:phase} shows the gas density--temperature phase diagram of Halo-11 and Halo-11-noSN. 
Due to the balance between Ly$\alpha$ cooling and radiative heating from UVB, 
a large fraction of gas results in neutral warm state with $T\sim10^{4}~\rm K$ in both runs. 
At $n_{\rm H} \gtrsim 1~\rm cm^{-3}$, the gas temperature gradually decreases to $<10^{4}~\rm K$ as density increases due to metal cooling. 
SN feedback suppresses the formation of high-density gas with $n_{\rm H} \gtrsim 10^{3}~\rm cm^{-3}$, but
some fraction of gas can reach $n_{\rm H} \sim 10^{3}-10^{4}~\rm cm^{-3}$. 
As shown in Equation~\ref{eq:nth}, SN feedback in this high-density region becomes inefficient due to radiative cooling. 

The apparent difference between Halo-11 and Halo-11noSN is the low-density, high-temperature region. 
Halo-11 shows some gas distribution at $n_{\rm H} \sim 10^{-3}-10^{-2}~\rm cm^{-3}$ and $T \sim 10^{5}-10^{6}~\rm K$. 
The low-density, high-temperature gas is shock-heated by SN feedback, and the temperature is corresponding to the relative velocity of $\sim 100~\rm km~s^{-1}$ between pre- and post-shock regions. 
For a low-metallicity gas, the cooling rate at $T\sim 10^{6}~\rm K$ is low, 
hence the shock-heated gas remains hot for $t_{\rm cool} \sim \frac{k_{\rm B}T}{n_{\rm H} \Lambda} \sim 2 \times 10^{7}~\rm yr$. 
Note that this time-scale is shorter than the typical dynamical time-scale of halos.
Therefore the hot gas can cool down and join star-forming regions as a warm medium if its outflowing velocity is less than the escape velocity of the halo.
In Halo-11noSN, there is no such low-density, high-temperature gas due to the lack of SN feedback. 
A small fraction of gas exists at $n_{\rm H} \sim 10^{-2}~\rm cm^{-3}$ and $T \sim 10^{5}-10^{6}~\rm K$ 
that is heated due to gravitational shocks. The gas cools down to a warm state after $t_{\rm cool} < 10^{7}~\rm yr$. 
The lower bounds are seen at high density $\nh \gtrsim 10^{2}~\rm cm^{-3}$. This is due to our effective EOS as explained in Sec.~\ref{sec:model} \citep[see also,][]{Johnson13, Schaye15}.
The slope of $4/3$ prevents spurious fragmentation due to the finite resolution \citep{Robertson08, DallaVecchia12}.

%Fig.11

\subsection{Baryonic impact on DM density profile}
\label{sec:dmprof}
It is well known that cosmological $N$-body simulations often show cusps in DM density profile at galactic centers nearly universally \citep[e.g.,][]{Navarro97, Fukushige97, Moore99, Jing00}.  
However, some observations of local galaxies indicated core-like structure of DM near the galactic center \citep[e.g.,][]{Gentile04}.
In order to solve the discrepancy, impacts of baryonic processes have been discussed \citep{Governato10, Governato12, Ogiya14, Onorbe15, Chan15}.
In particular, \citet{Pontzen12} showed that the rapid evacuation of gas from star-forming region near the galactic center 
results in the heating of DM and the formation of DM core. 
\citet{Davis14} showed that DM profiles of low-mass dwarf galaxies in the early Universe become shallower 
 due to SN feedback. 
As shown in the previous subsection, our simulations shows a rapid gas evacuation due to strong SN feedback. 
Therefore, the rapid change of baryonic distribution at the galactic center could also change the DM distribution. 

%Fig.12
\begin{figure}
\begin{center}
\includegraphics[scale=0.42]{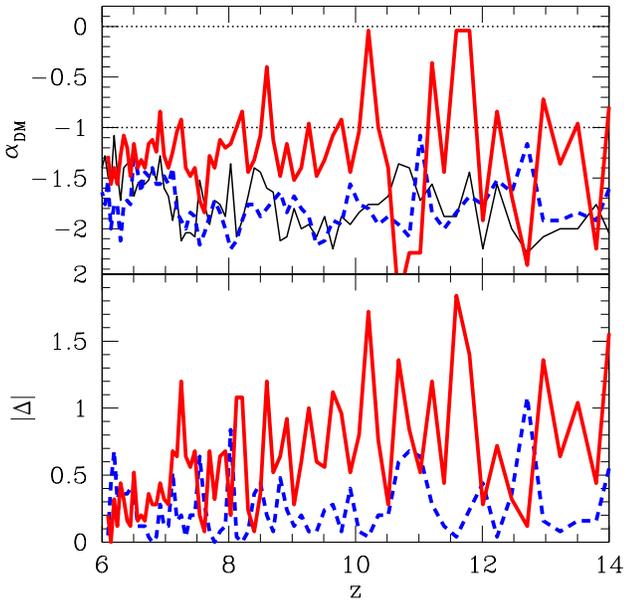}
\caption{
{\it Upper panel}: Redshift evolution of the power-law index of DM radial profile. 
Red solid, blue dash, and black solid lines represent Halo-11, Halo-11-lowSF, and Halo-11-noSN runs, respectively.  
{\it Lower panel}:  Absolute differences in the power-law indices between different runs. 
The red solid line shows the difference between Halo-11 and Halo-11-noSN, 
while blue dashed line is the one between Halo-11-lowSF and Halo-11-noSN.
}
\label{fig:dmprof}
\end{center}
\end{figure}

Figure~\ref{fig:dmprof} shows the redshift evolution of the power-law index of the dark matter density profile between $r=30 - 300~\rm pc$ in physical units.
The lower panel of the figure represents the differences of power-law indices between different runs.
We find that the DM profile of Halo-11 periodically changes, ranging from $\alpha_{\rm DM}\sim -1$ (cusp) to $\sim 0$ (core-like). 
After starbursts, DM profiles develop cores, but go back to cusps after a short time. 
The difference in $\alpha_{\rm DM}$ between Halo-11 and Halo-11-noSN frequently exceeds unity.
In the case of Halo-11-lowSF and Halo-11-noSN, the DM profile also changes with time, but it is always less than $\sim -1$. 
Therefore only the efficient SF model with SN feedback can produce cored profiles of DM. 
This oscillation of DM profile was also shown in previous works \citep[e.g.,][]{Pontzen12, Davis14}.
They claimed that the the oscillation of DM profile have a cumulative effect of heating up the DM, and keeps the core structure.  
On the other hand, \citet{Zhu16} 
found that DM cores were not seen in the dwarf galaxies in their moving mesh cosmological simulations.
Our current simulations also do not show a clear difference between simulations with and without SN feedback at $z \sim 6$.
Thus, we argue that the heating effect on DM up to $z \sim 6$ is not enough to make lasting cores. 

\citet{Davis14} showed the dependence of SF efficiency on the heating of DM.
They showed SFR $\gtrsim 0.2~\Msunyr$ was required to change the DM profile by SN feedback.
Our simulations indicate that the SF efficiency can be an important factor to the heating of DM
in the current SF and SN feedback models. 
The slow SF model tend to form stars in higher density regions, resulting in lower feedback efficiency owing to more efficient radiative cooling loss.  As a result, although the SFR of Halo-11-lowSFR is higher than Halo-11 as shown in Figure~\ref{fig:sfr}, the DM density profile does not change significantly with star formation. 

Note that, the current analysis is only one example.  
Furthermore, our simulations focus on high-redshift galaxies which are violently evolving 
via frequent merger events, whereas the observational data for the core profile are for well-relaxed, low-mass galaxies. 
 The relation between stellar feedback and DM density profile
should be investigated using a larger galaxy sample evolved to low redshifts.

%----------------------------------------------------------------------
%
% Section 4:  Discussion
%
%----------------------------------------------------------------------

%\section{Discussion} 
%\label{sec:discussion}

\subsection{External feedback by UV background radiation}

Star formation in first galaxies can also be affected by external UVB \citep[e.g.,][]{Susa04, Okamoto08a}.
In our fiducial model, we have turned on the UVB at $z = 10$ to mimic the effect of cosmic reionization. 
The photo-ionization by the UVB suppresses the cooling rate of primordial gas at $T \sim 10^{4}-10^{5}$ K.
Recent observations of the cosmic microwave background have indicated that the cosmic reionization 
took place at $z \sim 9 - 11$ \citep{Komatsu11, Planck16}.
Cosmic reionization proceeds with a patchy ionization structure as shown in the work using cosmological simulations \citep[e.g.,][]{Iliev14}.
By $z \sim 6$, cosmic reionization is completed once the H{\sc ii} bubbles completely overlap. 
This patchy reionization scenario suggests that the external UVB sensitively depends on the environment.
If galaxies reside in high-density regions where the ionization of IGM proceeds earlier, 
they are likely to be affected by the external UV flux at higher redshift. 
Here we investigate the impact of different reionization redshifts, when the UVB starts to irradiate, on the SF history of galaxies, as one of environmental effects. 
Using the same initial condition as Halo-11, we carry out an additional simulation
with the reionization redshift $z_{\rm re}=18$ (Halo-11-eUVB).

\begin{figure}
\begin{center}
\includegraphics[scale=0.4]{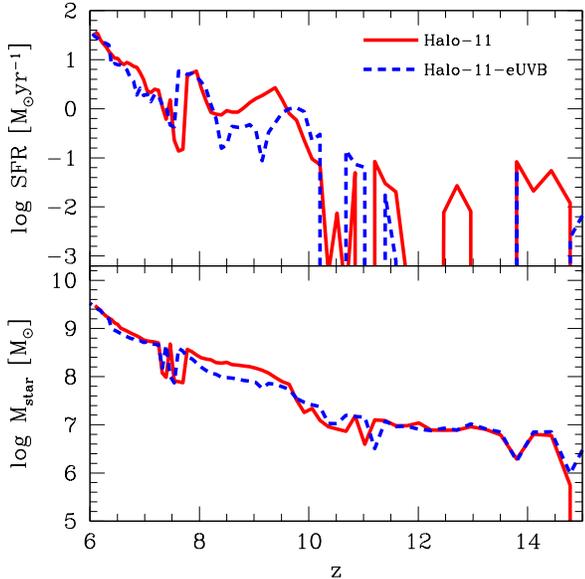}
\caption{
SFR and stellar mass assembly history for simulations with different
reionization redshifts $z_{\rm re}=10$ (Halo-11) and $z_{\rm re}=18$ (Halo-11-eUVB). 
}
\label{fig:sfr_uv}
\end{center}
\end{figure}

Figure~\ref{fig:sfr_uv} shows SFR and stellar mass assembly histories. 
A clear difference is seen in the SFR at $z > 10$. 
The SFR of both simulations proceeds intermittently due to the SN feedback
as explained in Section~\ref{fig:sfr}.
However, the quenching time of star formation is longer in the case of Halo-11-eUVB.
Since the mean gas density in halos at $z > 6$ is $n_{\rm H} \gtrsim 10^{-2}~\rm cm^{-3}$, 
the interstellar medium in high-redshift galaxies is easily self-shielded from UVB. 
However, once the interstellar gas expands by the SN feedback, 
the decreased gas density allows the UVB to penetrate and ionize the gas.  
As a result, the photoionization heating decreases the gas density further and suppresses star formation. 
This make the no-SF phase longer in Halo-11-eUVB at $z > 10$. 
At $z \le 10$, the UVB is turned on in both simulations. 
The different SF history due to UVB continues until $z \sim 8$, and then the SFR  converges at $z \lesssim 8$. 
In addition, the photo-ionized gas by the UVB with $T \sim 10^{4}~\rm K$ 
can be bound in the gravitation potential of halos at $z \lesssim 10$.
Thus, the impact of UVB on SF history can be secondary compared to the SN feedback at lower redshifts.

Unlike the SFR history, the stellar mass assembly history under different UVB models does not show clear differences. 
The low SFR phase does not contribute to the stellar mass assembly.
The high SFR phase does not depend on the UVB models very much, because
 the star-forming clouds is completely shielded from the UVB. 
As a result, the difference in the stellar mass is always within twenty per cent. 
Thus we suggest that the environmental effect of UVB on galaxy formation is not so strong in our galaxy sample. 
Note that, however, the large dispersion of SFR can change the shape of the faint-end slope of UV luminosity function. 
Hence, combining UV luminosity and stellar mass function allows us to understand the SF history and 
feedback efficiency.  The current work focuses on a small region with a high resolution, hence we cannot discuss the statistical feature of the first galaxies.  
We will investigate the statistical feature using a larger galaxy sample and multi-wavelength radiative transfer calculations in our future work. 

%----------------------------------------------------------------------
%
% Section 5:  Summary
%
%----------------------------------------------------------------------

\section{Discussion and Summary}
\label{sec:summary}
We investigate the formation of the first galaxies and their evolution to $z \sim 6$
using high-resolution cosmological hydrodynamic simulations with zoom-in initial conditions.
We focus on three different galaxies in halos with masses 
of $\Mh \sim10^{10}~\Msun$ (Halo-10), $\sim 10^{11}~\Msun$ (Halo-11) and $\sim 10^{12}~\Msun$ at $z=6$.
In particular, we study the effects of SN feedback and different SF efficiencies on 
the physical properties of the first galaxies. 
We also examined the impact of UVB as an external feedback, motivated by the patchy reionization scenario. 

Our major findings are as follows: 

\begin{enumerate}
\item
Gas in star-forming regions is expelled out of the galactic potential due to SN feedback, and star formation is quenched.  As a result, the SF history becomes very intermittent as a function of time at $z \gtrsim 10$ when the halo mass is less than $\sim 10^{10}~\Msun$. 
At $z \lesssim 10$, the halo mass exceeds $\sim 10^{10}~\Msun$, and 
the gas can be held in galaxies against stellar feedback due to deeper gravitation potential wells. This allows galaxies to form stars continuously. 
In addition, the feature of intermittent star formation disappears in the SF model with a lower SF efficiency.
\item
SN feedback causes a very clumpy gas distribution, and prohibits the formation of galactic disk at $z>6$ by continuously disturbing an organized flow of gas.  
When we reduce the SF efficiency or turn off the SN feedback, 
galaxies can preserve a gaseous disk for a long period of time.
\item
Due to the expulsion of gas from the galactic center by SN feedback,  the stellar distribution becomes more extended compared to the case without SN feedback. This is because SN feedback suppresses the concentration of gas near the galactic center, whereas in the case with no SN feedback stars efficiently form in high-density clouds near the galactic center.  As a result, the half-mass radius of stars can become greater than $\sim 0.1 R_{\rm vir}$ owing to SN feedback. 
\item
SN feedback can change the density profile of dark matter in the galactic center, producing a core for a short period of time after the starburst.  
However, the frequent changes of gravitational potential due to SN feedback does not cause any clear differences between the runs with and without SN feedback at $z\sim 6$.
This suggests that the cumulative effect of SN feedback till $z\sim 6$ is insufficient to sustain a cored DM profile for a long period of time. 
\item
The galaxies in Halo-11 and Halo-12 reproduce the SFR of observed LAEs with $\sim 10~\Msunyr$ \citep[e.g.,][]{Shibuya12}.
The SFR of the massive galaxy in Halo-12 becomes higher than $10~\Msunyr$ at $z \lesssim 8$, which corresponds to the level observed for bright LBGs \citep[e.g.,][]{Bouwens15}. 
\item
In the patchy reionization scenario, galaxies residing in overdense regions are likely to be irradiated by the external UV radiation earlier.  We examine the effect of UVB on the SF history by changing the reionization redshift when the UVB turns on.  We find that the UVB extends the suppression time of star formation caused by the SN feedback at $z \gtrsim 10$, 
but does not affect the total cumulative stellar mass at $z=6$.
Therefore, the impact of UVB on the our galaxy star formation is secondary compared to SN feedback.
\end{enumerate}

In this work, by considering the wide halo mass range up to $\Mh \sim 10^{12}~\Msun$ using zoom-in initial conditions, we directly compare massive modeled galaxies with various observed galaxies, e.g., LAEs, LBGs, and SMGs at $z \gtrsim 6$,
whereas some previous works were limited with $\Mh \lesssim 10^{11}~\Msun$ \citep[e.g.,][]{Johnson13, Hopkins14}. In addition, the wide halo mass range allow us to follow the transition of star formation mode from the intermittent star formation history to the stable continuous star formation. 

The gas expulsion from the galactic center probably affects the co-evolution of central massive black hole (BH) and the stellar bulge. 
\citet{Dubois15} recently demonstrated that the SN feedback can eject gas from galaxies, and significantly suppress the growth of  a central BH \citep[e.g.,][]{Hopkins16}. 
Our simulations indicate that the gas near the galactic center has low densities, and suggest slower growth of BHs for those galaxies that have been affected strongly by SN feedback. 

In addition, the gas outflow from star-forming regions can allow easier escape of ionizing photons from young stars 
which caused cosmic reionization \citep[e.g.,][]{Yajima11, Yajima14c, Paardekooper15, Ma15, Kimm14, Kimm17}.

Our results highlight the intermittent SF activities caused by the SN feedback even in massive halos. 
However, the sample size of simulated galaxies in the present work is limited due to the zoom-in technique that we employed to achieve a higher resolution than full-box cosmological simulations. 
Therefore, a statistical study of the nature of the first galaxies is beyond our scope in this work.  
In the future, we will present a comprehensive study of a large sample of zoom-in simulations with comparable numerical resolution, and investigate the statical nature of first galaxies, such as luminosity function, stellar mass function, clustering, and variance in escape fraction of ionizing photons 
for upcoming large telescopes (JWST, TMT, E-ELT). 
 
%----------------------------------------------------------------------
%
% Acknowledge
%
%----------------------------------------------------------------------
\section*{Acknowledgments}
The numerical simulations were performed on the computer cluster, {\tt Draco}, at the Frontier Research Institute for Interdisciplinary Sciences of Tohoku University and {\tt Pluto} in the Theoretical Astrophysics Group at Osaka University.  This work is supported in part by the MEXT/JSPS KAKENHI Grant Number 15H06022 (HY) and JP26247022 (KN).
CDV acknowledges financial support from the Spanish Ministry of Economy and Competitiveness (MINECO) under the 2015 Severo Ochoa program SEV-2015-0548, and grants AYA2014-58308 and RYC-2015-18078. 

%----------------------------------------------------------------------
%
% References
%
%----------------------------------------------------------------------
\bibliographystyle{apj}

\begin{thebibliography}{90}
\expandafter\ifx\csname natexlab\endcsname\relax\def\natexlab#1{#1}\fi

\bibitem[{{Agarwal} \& {Khochfar}(2015)}]{Agarwal15}
{Agarwal}, B., \& {Khochfar}, S. 2015, \mnras, 446, 160

\bibitem[{{Behroozi} {et~al.}(2013){Behroozi}, {Wechsler}, \&
  {Conroy}}]{Behroozi13}
{Behroozi}, P.~S., {Wechsler}, R.~H., \& {Conroy}, C. 2013, \apj, 770, 57

\bibitem[{{Bouwens} {et~al.}(2009){Bouwens}, {Illingworth}, {Franx}, {Chary},
  {Meurer}, {Conselice}, {Ford}, {Giavalisco}, \& {van Dokkum}}]{Bouwens09}
{Bouwens}, R.~J., {Illingworth}, G.~D., {Franx}, M., {Chary}, R.-R., {Meurer},
  G.~R., {Conselice}, C.~J., {Ford}, H., {Giavalisco}, M., \& {van Dokkum}, P.
  2009, \apj, 705, 936

\bibitem[{{Bouwens} {et~al.}(2015){Bouwens}, {Illingworth}, {Oesch}, {Trenti},
  {Labb{\'e}}, {Bradley}, {Carollo}, {van Dokkum}, {Gonzalez}, {Holwerda},
  {Franx}, {Spitler}, {Smit}, \& {Magee}}]{Bouwens15}
{Bouwens}, R.~J., {Illingworth}, G.~D., {Oesch}, P.~A., {Trenti}, M.,
  {Labb{\'e}}, I., {Bradley}, L., {Carollo}, M., {van Dokkum}, P.~G.,
  {Gonzalez}, V., {Holwerda}, B., {Franx}, M., {Spitler}, L., {Smit}, R., \&
  {Magee}, D. 2015, \apj, 803, 34

\bibitem[{{Bouwens} {et~al.}(2010){Bouwens}, {Illingworth}, {Oesch}, {Trenti},
  {Stiavelli}, {Carollo}, {Franx}, {van Dokkum}, {Labb{\'e}}, \&
  {Magee}}]{Bouwens10}
{Bouwens}, R.~J., {Illingworth}, G.~D., {Oesch}, P.~A., {Trenti}, M.,
  {Stiavelli}, M., {Carollo}, C.~M., {Franx}, M., {van Dokkum}, P.~G.,
  {Labb{\'e}}, I., \& {Magee}, D. 2010, \apjl, 708, L69

\bibitem[{{Bowler} {et~al.}(2015){Bowler}, {Dunlop}, {McLure}, {McCracken},
  {Milvang-Jensen}, {Furusawa}, {Taniguchi}, {Le F{\`e}vre}, {Fynbo}, {Jarvis},
  \& {H{\"a}u{\ss}ler}}]{Bowler15}
{Bowler}, R.~A.~A., {Dunlop}, J.~S., {McLure}, R.~J., {McCracken}, H.~J.,
  {Milvang-Jensen}, B., {Furusawa}, H., {Taniguchi}, Y., {Le F{\`e}vre}, O.,
  {Fynbo}, J.~P.~U., {Jarvis}, M.~J., \& {H{\"a}u{\ss}ler}, B. 2015, \mnras,
  452, 1817

\bibitem[{{Chan} {et~al.}(2015){Chan}, {Kere{\v s}}, {O{\~n}orbe}, {Hopkins},
  {Muratov}, {Faucher-Gigu{\`e}re}, \& {Quataert}}]{Chan15}
{Chan}, T.~K., {Kere{\v s}}, D., {O{\~n}orbe}, J., {Hopkins}, P.~F., {Muratov},
  A.~L., {Faucher-Gigu{\`e}re}, C.-A., \& {Quataert}, E. 2015, \mnras, 454,
  2981

\bibitem[{{Cioffi} {et~al.}(1988){Cioffi}, {McKee}, \&
  {Bertschinger}}]{Cioffi88}
{Cioffi}, D.~F., {McKee}, C.~F., \& {Bertschinger}, E. 1988, \apj, 334, 252

\bibitem[{{Cullen} {et~al.}(2017){Cullen}, {McLure}, {Khochfar}, {Dunlop}, \&
  {Dalla Vecchia}}]{Cullen17}
{Cullen}, F., {McLure}, R.~J., {Khochfar}, S., {Dunlop}, J.~S., \& {Dalla
  Vecchia}, C. 2017, ArXiv e-prints

\bibitem[{{Daddi} {et~al.}(2007){Daddi}, {Dickinson}, {Morrison}, {Chary},
  {Cimatti}, {Elbaz}, {Frayer}, {Renzini}, {Pope}, {Alexander}, {Bauer},
  {Giavalisco}, {Huynh}, {Kurk}, \& {Mignoli}}]{Daddi07}
{Daddi}, E., {Dickinson}, M., {Morrison}, G., {Chary}, R., {Cimatti}, A.,
  {Elbaz}, D., {Frayer}, D., {Renzini}, A., {Pope}, A., {Alexander}, D.~M.,
  {Bauer}, F.~E., {Giavalisco}, M., {Huynh}, M., {Kurk}, J., \& {Mignoli}, M.
  2007, \apj, 670, 156

\bibitem[{{Dalla Vecchia} \& {Schaye}(2012)}]{DallaVecchia12}
{Dalla Vecchia}, C., \& {Schaye}, J. 2012, \mnras, 426, 140

\bibitem[{{Davis} {et~al.}(2014){Davis}, {Khochfar}, \& {Dalla
  Vecchia}}]{Davis14}
{Davis}, A.~J., {Khochfar}, S., \& {Dalla Vecchia}, C. 2014, \mnras, 443, 985

\bibitem[{{Dubois} {et~al.}(2015){Dubois}, {Volonteri}, {Silk}, {Devriendt},
  {Slyz}, \& {Teyssier}}]{Dubois15}
{Dubois}, Y., {Volonteri}, M., {Silk}, J., {Devriendt}, J., {Slyz}, A., \&
  {Teyssier}, R. 2015, \mnras, 452, 1502

\bibitem[{{Elliott} {et~al.}(2015){Elliott}, {Khochfar}, {Greiner}, \& {Dalla
  Vecchia}}]{Elliot15}
{Elliott}, J., {Khochfar}, S., {Greiner}, J., \& {Dalla Vecchia}, C. 2015,
  \mnras, 446, 4239

\bibitem[{{Ferland}(2000)}]{Ferland00}
{Ferland}, G.~J. 2000, in Revista Mexicana de Astronomia y Astrofisica
  Conference Series, Vol.~9, Revista Mexicana de Astronomia y Astrofisica
  Conference Series, ed. S.~J. {Arthur}, N.~S. {Brickhouse}, \& J.~{Franco},
  153--157

\bibitem[{{Finkelstein} {et~al.}(2011){Finkelstein}, {Hill}, {Gebhardt},
  {Adams}, {Blanc}, {Papovich}, {Ciardullo}, {Drory}, {Gawiser}, {Gronwall},
  {Schneider}, \& {Tran}}]{Finkelstein11}
{Finkelstein}, S.~L., {Hill}, G.~J., {Gebhardt}, K., {Adams}, J., {Blanc},
  G.~A., {Papovich}, C., {Ciardullo}, R., {Drory}, N., {Gawiser}, E.,
  {Gronwall}, C., {Schneider}, D.~P., \& {Tran}, K.-V. 2011, \apj, 729, 140

\bibitem[{{Finkelstein} {et~al.}(2013){Finkelstein}, {Papovich}, {Dickinson},
  {Song}, {Tilvi}, {Koekemoer}, {Finkelstein}, {Mobasher}, {Ferguson},
  {Giavalisco}, {Reddy}, {Ashby}, {Dekel}, {Fazio}, {Fontana}, {Grogin},
  {Huang}, {Kocevski}, {Rafelski}, {Weiner}, \& {Willner}}]{Finkelstein13}
{Finkelstein}, S.~L., {Papovich}, C., {Dickinson}, M., {Song}, M., {Tilvi}, V.,
  {Koekemoer}, A.~M., {Finkelstein}, K.~D., {Mobasher}, B., {Ferguson}, H.~C.,
  {Giavalisco}, M., {Reddy}, N., {Ashby}, M.~L.~N., {Dekel}, A., {Fazio},
  G.~G., {Fontana}, A., {Grogin}, N.~A., {Huang}, J.-S., {Kocevski}, D.,
  {Rafelski}, M., {Weiner}, B.~J., \& {Willner}, S.~P. 2013, \nat, 502, 524

\bibitem[{{Fukushige} \& {Makino}(1997)}]{Fukushige97}
{Fukushige}, T., \& {Makino}, J. 1997, \apjl, 477, L9

\bibitem[{{Gentile} {et~al.}(2004){Gentile}, {Salucci}, {Klein}, {Vergani}, \&
  {Kalberla}}]{Gentile04}
{Gentile}, G., {Salucci}, P., {Klein}, U., {Vergani}, D., \& {Kalberla}, P.
  2004, \mnras, 351, 903

\bibitem[{{Genzel} {et~al.}(2010){Genzel}, {Tacconi}, {Gracia-Carpio},
  {Sternberg}, {Cooper}, {Shapiro}, {Bolatto}, {Bouch{\'e}}, {Bournaud},
  {Burkert}, {Combes}, {Comerford}, {Cox}, {Davis}, {Schreiber},
  {Garcia-Burillo}, {Lutz}, {Naab}, {Neri}, {Omont}, {Shapley}, \&
  {Weiner}}]{Genzel10}
{Genzel}, R., {Tacconi}, L.~J., {Gracia-Carpio}, J., {Sternberg}, A., {Cooper},
  M.~C., {Shapiro}, K., {Bolatto}, A., {Bouch{\'e}}, N., {Bournaud}, F.,
  {Burkert}, A., {Combes}, F., {Comerford}, J., {Cox}, P., {Davis}, M.,
  {Schreiber}, N.~M.~F., {Garcia-Burillo}, S., {Lutz}, D., {Naab}, T., {Neri},
  R., {Omont}, A., {Shapley}, A., \& {Weiner}, B. 2010, \mnras, 407, 2091

\bibitem[{{Governato} {et~al.}(2010){Governato}, {Brook}, {Mayer}, {Brooks},
  {Rhee}, {Wadsley}, {Jonsson}, {Willman}, {Stinson}, {Quinn}, \&
  {Madau}}]{Governato10}
{Governato}, F., {Brook}, C., {Mayer}, L., {Brooks}, A., {Rhee}, G., {Wadsley},
  J., {Jonsson}, P., {Willman}, B., {Stinson}, G., {Quinn}, T., \& {Madau}, P.
  2010, \nat, 463, 203

\bibitem[{{Governato} {et~al.}(2012){Governato}, {Zolotov}, {Pontzen},
  {Christensen}, {Oh}, {Brooks}, {Quinn}, {Shen}, \& {Wadsley}}]{Governato12}
{Governato}, F., {Zolotov}, A., {Pontzen}, A., {Christensen}, C., {Oh}, S.~H.,
  {Brooks}, A.~M., {Quinn}, T., {Shen}, S., \& {Wadsley}, J. 2012, \mnras, 422,
  1231

\bibitem[{{Haardt} \& {Madau}(2001)}]{Haardt01}
{Haardt}, F., \& {Madau}, P. 2001, in Clusters of Galaxies and the High
  Redshift Universe Observed in X-rays, ed. D.~M. {Neumann} \& J.~T.~V. {Tran}

\bibitem[{{Hahn} \& {Abel}(2011)}]{Hahn11}
{Hahn}, O., \& {Abel}, T. 2011, \mnras, 415, 2101

\bibitem[{{Hasegawa} \& {Semelin}(2013)}]{Hasegawa13}
{Hasegawa}, K., \& {Semelin}, B. 2013, \mnras, 428, 154

\bibitem[{{Hopkins} {et~al.}(2014){Hopkins}, {Kere{\v s}}, {O{\~n}orbe},
  {Faucher-Gigu{\`e}re}, {Quataert}, {Murray}, \& {Bullock}}]{Hopkins14}
{Hopkins}, P.~F., {Kere{\v s}}, D., {O{\~n}orbe}, J., {Faucher-Gigu{\`e}re},
  C.-A., {Quataert}, E., {Murray}, N., \& {Bullock}, J.~S. 2014, \mnras, 445,
  581

\bibitem[{{Hopkins} {et~al.}(2016){Hopkins}, {Torrey}, {Faucher-Gigu{\`e}re},
  {Quataert}, \& {Murray}}]{Hopkins16}
{Hopkins}, P.~F., {Torrey}, P., {Faucher-Gigu{\`e}re}, C.-A., {Quataert}, E.,
  \& {Murray}, N. 2016, \mnras, 458, 816

\bibitem[{{Iliev} {et~al.}(2014){Iliev}, {Mellema}, {Ahn}, {Shapiro}, {Mao}, \&
  {Pen}}]{Iliev14}
{Iliev}, I.~T., {Mellema}, G., {Ahn}, K., {Shapiro}, P.~R., {Mao}, Y., \&
  {Pen}, U.-L. 2014, \mnras, 439, 725

\bibitem[{{Iye} {et~al.}(2006){Iye}, {Ota}, {Kashikawa}, {Furusawa},
  {Hashimoto}, {Hattori}, {Matsuda}, {Morokuma}, {Ouchi}, \&
  {Shimasaku}}]{Iye06}
{Iye}, M., {Ota}, K., {Kashikawa}, N., {Furusawa}, H., {Hashimoto}, T.,
  {Hattori}, T., {Matsuda}, Y., {Morokuma}, T., {Ouchi}, M., \& {Shimasaku}, K.
  2006, \nat, 443, 186

\bibitem[{{Jeon} {et~al.}(2014){Jeon}, {Pawlik}, {Bromm}, \&
  {Milosavljevi{\'c}}}]{Jeon14}
{Jeon}, M., {Pawlik}, A.~H., {Bromm}, V., \& {Milosavljevi{\'c}}, M. 2014,
  \mnras, 440, 3778

\bibitem[{{Jing} \& {Suto}(2000)}]{Jing00}
{Jing}, Y.~P., \& {Suto}, Y. 2000, \apjl, 529, L69

\bibitem[{{Johnson} {et~al.}(2013){Johnson}, {Dalla}, \&
  {Khochfar}}]{Johnson13}
{Johnson}, J.~L., {Dalla}, V.~C., \& {Khochfar}, S. 2013, \mnras, 428, 1857

\bibitem[{{Kennicutt}(1998)}]{Kennicutt98}
{Kennicutt}, Jr., R.~C. 1998, \araa, 36, 189

\bibitem[{{Kim} \& {Ostriker}(2015)}]{Kim15}
{Kim}, C.-G., \& {Ostriker}, E.~C. 2015, \apj, 802, 99

\bibitem[{{Kimm} \& {Cen}(2014)}]{Kimm14}
{Kimm}, T., \& {Cen}, R. 2014, \apj, 788, 121

\bibitem[{{Kimm} {et~al.}(2017){Kimm}, {Katz}, {Haehnelt}, {Rosdahl},
  {Devriendt}, \& {Slyz}}]{Kimm17}
{Kimm}, T., {Katz}, H., {Haehnelt}, M., {Rosdahl}, J., {Devriendt}, J., \&
  {Slyz}, A. 2017, \mnras

\bibitem[{{Komatsu} {et~al.}(2011){Komatsu}, {Smith}, {Dunkley}, {Bennett},
  {Gold}, {Hinshaw}, {Jarosik}, {Larson}, {Nolta}, {Page}, {Spergel},
  {Halpern}, {Hill}, {Kogut}, {Limon}, {Meyer}, {Odegard}, {Tucker}, {Weiland},
  {Wollack}, \& {Wright}}]{Komatsu11}
{Komatsu}, E., {Smith}, K.~M., {Dunkley}, J., {Bennett}, C.~L., {Gold}, B.,
  {Hinshaw}, G., {Jarosik}, N., {Larson}, D., {Nolta}, M.~R., {Page}, L.,
  {Spergel}, D.~N., {Halpern}, M., {Hill}, R.~S., {Kogut}, A., {Limon}, M.,
  {Meyer}, S.~S., {Odegard}, N., {Tucker}, G.~S., {Weiland}, J.~L., {Wollack},
  E., \& {Wright}, E.~L. 2011, \apjs, 192, 18

\bibitem[{{Konno} {et~al.}(2014){Konno}, {Ouchi}, {Ono}, {Shimasaku},
  {Shibuya}, {Furusawa}, {Nakajima}, {Naito}, {Momose}, {Yuma}, \&
  {Iye}}]{Konno14}
{Konno}, A., {Ouchi}, M., {Ono}, Y., {Shimasaku}, K., {Shibuya}, T.,
  {Furusawa}, H., {Nakajima}, K., {Naito}, Y., {Momose}, R., {Yuma}, S., \&
  {Iye}, M. 2014, \apj, 797, 16

\bibitem[{{Ma} {et~al.}(2015){Ma}, {Kasen}, {Hopkins}, {Faucher-Gigu{\`e}re},
  {Quataert}, {Kere{\v s}}, \& {Murray}}]{Ma15}
{Ma}, X., {Kasen}, D., {Hopkins}, P.~F., {Faucher-Gigu{\`e}re}, C.-A.,
  {Quataert}, E., {Kere{\v s}}, D., \& {Murray}, N. 2015, \mnras, 453, 960

\bibitem[{{Madau} {et~al.}(1998){Madau}, {Pozzetti}, \& {Dickinson}}]{Madau98}
{Madau}, P., {Pozzetti}, L., \& {Dickinson}, M. 1998, \apj, 498, 106

\bibitem[{{M{\'a}rmol-Queralt{\'o}} {et~al.}(2016){M{\'a}rmol-Queralt{\'o}},
  {McLure}, {Cullen}, {Dunlop}, {Fontana}, \& {McLeod}}]{Marmol16}
{M{\'a}rmol-Queralt{\'o}}, E., {McLure}, R.~J., {Cullen}, F., {Dunlop}, J.~S.,
  {Fontana}, A., \& {McLeod}, D.~J. 2016, \mnras, 460, 3587

\bibitem[{{McLure} {et~al.}(2009){McLure}, {Cirasuolo}, {Dunlop}, {Foucaud}, \&
  {Almaini}}]{McLure09}
{McLure}, R.~J., {Cirasuolo}, M., {Dunlop}, J.~S., {Foucaud}, S., \& {Almaini},
  O. 2009, \mnras, 395, 2196

\bibitem[{{McLure} {et~al.}(2013){McLure}, {Dunlop}, {Bowler}, {Curtis-Lake},
  {Schenker}, {Ellis}, {Robertson}, {Koekemoer}, {Rogers}, {Ono}, {Ouchi},
  {Charlot}, {Wild}, {Stark}, {Furlanetto}, {Cirasuolo}, \&
  {Targett}}]{McLure13}
{McLure}, R.~J., {Dunlop}, J.~S., {Bowler}, R.~A.~A., {Curtis-Lake}, E.,
  {Schenker}, M., {Ellis}, R.~S., {Robertson}, B.~E., {Koekemoer}, A.~M.,
  {Rogers}, A.~B., {Ono}, Y., {Ouchi}, M., {Charlot}, S., {Wild}, V., {Stark},
  D.~P., {Furlanetto}, S.~R., {Cirasuolo}, M., \& {Targett}, T.~A. 2013,
  \mnras, 432, 2696

\bibitem[{{Mo} {et~al.}(1998){Mo}, {Mao}, \& {White}}]{Mo98}
{Mo}, H.~J., {Mao}, S., \& {White}, S.~D.~M. 1998, \mnras, 295, 319

\bibitem[{{Moore} {et~al.}(1999){Moore}, {Ghigna}, {Governato}, {Lake},
  {Quinn}, {Stadel}, \& {Tozzi}}]{Moore99}
{Moore}, B., {Ghigna}, S., {Governato}, F., {Lake}, G., {Quinn}, T., {Stadel},
  J., \& {Tozzi}, P. 1999, \apjl, 524, L19

\bibitem[{{Nagamine} {et~al.}(2010){Nagamine}, {Choi}, \&
  {Yajima}}]{Nagamine10a}
{Nagamine}, K., {Choi}, J., \& {Yajima}, H. 2010, \apjl, 725, L219

\bibitem[{{Navarro} {et~al.}(1997){Navarro}, {Frenk}, \& {White}}]{Navarro97}
{Navarro}, J.~F., {Frenk}, C.~S., \& {White}, S.~D.~M. 1997, \apj, 490, 493

\bibitem[{{O{\~n}orbe} {et~al.}(2015){O{\~n}orbe}, {Boylan-Kolchin}, {Bullock},
  {Hopkins}, {Kere{\v s}}, {Faucher-Gigu{\`e}re}, {Quataert}, \&
  {Murray}}]{Onorbe15}
{O{\~n}orbe}, J., {Boylan-Kolchin}, M., {Bullock}, J.~S., {Hopkins}, P.~F.,
  {Kere{\v s}}, D., {Faucher-Gigu{\`e}re}, C.-A., {Quataert}, E., \& {Murray},
  N. 2015, \mnras, 454, 2092

\bibitem[{{Oesch} {et~al.}(2013){Oesch}, {Bouwens}, {Illingworth}, {Labb{\'e}},
  {Franx}, {van Dokkum}, {Trenti}, {Stiavelli}, {Gonzalez}, \&
  {Magee}}]{Oesch13}
{Oesch}, P.~A., {Bouwens}, R.~J., {Illingworth}, G.~D., {Labb{\'e}}, I.,
  {Franx}, M., {van Dokkum}, P.~G., {Trenti}, M., {Stiavelli}, M., {Gonzalez},
  V., \& {Magee}, D. 2013, \apj, 773, 75

\bibitem[{{Oesch} {et~al.}(2014){Oesch}, {Bouwens}, {Illingworth}, {Labb{\'e}},
  {Smit}, {Franx}, {van Dokkum}, {Momcheva}, {Ashby}, {Fazio}, {Huang},
  {Willner}, {Gonzalez}, {Magee}, {Trenti}, {Brammer}, {Skelton}, \&
  {Spitler}}]{Oesch14}
{Oesch}, P.~A., {Bouwens}, R.~J., {Illingworth}, G.~D., {Labb{\'e}}, I.,
  {Smit}, R., {Franx}, M., {van Dokkum}, P.~G., {Momcheva}, I., {Ashby},
  M.~L.~N., {Fazio}, G.~G., {Huang}, J.-S., {Willner}, S.~P., {Gonzalez}, V.,
  {Magee}, D., {Trenti}, M., {Brammer}, G.~B., {Skelton}, R.~E., \& {Spitler},
  L.~R. 2014, \apj, 786, 108

\bibitem[{{Oesch} {et~al.}(2016){Oesch}, {Brammer}, {van Dokkum},
  {Illingworth}, {Bouwens}, {Labb{\'e}}, {Franx}, {Momcheva}, {Ashby}, {Fazio},
  {Gonzalez}, {Holden}, {Magee}, {Skelton}, {Smit}, {Spitler}, {Trenti}, \&
  {Willner}}]{Oesch16}
{Oesch}, P.~A., {Brammer}, G., {van Dokkum}, P.~G., {Illingworth}, G.~D.,
  {Bouwens}, R.~J., {Labb{\'e}}, I., {Franx}, M., {Momcheva}, I., {Ashby},
  M.~L.~N., {Fazio}, G.~G., {Gonzalez}, V., {Holden}, B., {Magee}, D.,
  {Skelton}, R.~E., {Smit}, R., {Spitler}, L.~R., {Trenti}, M., \& {Willner},
  S.~P. 2016, \apj, 819, 129

\bibitem[{{Oesch} {et~al.}(2015){Oesch}, {van Dokkum}, {Illingworth},
  {Bouwens}, {Momcheva}, {Holden}, {Roberts-Borsani}, {Smit}, {Franx},
  {Labb{\'e}}, {Gonz{\'a}lez}, \& {Magee}}]{Oesch15}
{Oesch}, P.~A., {van Dokkum}, P.~G., {Illingworth}, G.~D., {Bouwens}, R.~J.,
  {Momcheva}, I., {Holden}, B., {Roberts-Borsani}, G.~W., {Smit}, R., {Franx},
  M., {Labb{\'e}}, I., {Gonz{\'a}lez}, V., \& {Magee}, D. 2015, \apjl, 804, L30

\bibitem[{{Ogiya} \& {Mori}(2014)}]{Ogiya14}
{Ogiya}, G., \& {Mori}, M. 2014, \apj, 793, 46

\bibitem[{{Okamoto} {et~al.}(2008){Okamoto}, {Gao}, \& {Theuns}}]{Okamoto08a}
{Okamoto}, T., {Gao}, L., \& {Theuns}, T. 2008, MNRAS, 390, 920

\bibitem[{{Ono} {et~al.}(2012){Ono}, {Ouchi}, {Mobasher}, {Dickinson},
  {Penner}, {Shimasaku}, {Weiner}, {Kartaltepe}, {Nakajima}, {Nayyeri},
  {Stern}, {Kashikawa}, \& {Spinrad}}]{Ono12}
{Ono}, Y., {Ouchi}, M., {Mobasher}, B., {Dickinson}, M., {Penner}, K.,
  {Shimasaku}, K., {Weiner}, B.~J., {Kartaltepe}, J.~S., {Nakajima}, K.,
  {Nayyeri}, H., {Stern}, D., {Kashikawa}, N., \& {Spinrad}, H. 2012, \apj,
  744, 83

\bibitem[{{Ouchi} {et~al.}(2009){Ouchi}, {Mobasher}, {Shimasaku}, {Ferguson},
  {Fall}, {Ono}, {Kashikawa}, {Morokuma}, {Nakajima}, {Okamura}, {Dickinson},
  {Giavalisco}, \& {Ohta}}]{Ouchi09b}
{Ouchi}, M., {Mobasher}, B., {Shimasaku}, K., {Ferguson}, H.~C., {Fall}, S.~M.,
  {Ono}, Y., {Kashikawa}, N., {Morokuma}, T., {Nakajima}, K., {Okamura}, S.,
  {Dickinson}, M., {Giavalisco}, M., \& {Ohta}, K. 2009, \apj, 706, 1136

\bibitem[{{Ouchi} {et~al.}(2010){Ouchi}, {Shimasaku}, {Furusawa}, {Saito},
  {Yoshida}, {Akiyama}, {Ono}, {Yamada}, {Ota}, {Kashikawa}, {Iye}, {Kodama},
  {Okamura}, {Simpson}, \& {Yoshida}}]{Ouchi10}
{Ouchi}, M., {Shimasaku}, K., {Furusawa}, H., {Saito}, T., {Yoshida}, M.,
  {Akiyama}, M., {Ono}, Y., {Yamada}, T., {Ota}, K., {Kashikawa}, N., {Iye},
  M., {Kodama}, T., {Okamura}, S., {Simpson}, C., \& {Yoshida}, M. 2010, \apj,
  723, 869

\bibitem[{{Paardekooper} {et~al.}(2013){Paardekooper}, {Khochfar}, \& {Dalla
  Vecchia}}]{Paardekooper13}
{Paardekooper}, J.-P., {Khochfar}, S., \& {Dalla Vecchia}, C. 2013, \mnras,
  429, L94

\bibitem[{{Paardekooper} {et~al.}(2015){Paardekooper}, {Khochfar}, \& {Dalla
  Vecchia}}]{Paardekooper15}
---. 2015, \mnras, 451, 2544

\bibitem[{{Pawlik} {et~al.}(2011){Pawlik}, {Milosavljevi{\'c}}, \&
  {Bromm}}]{Pawlik11}
{Pawlik}, A.~H., {Milosavljevi{\'c}}, M., \& {Bromm}, V. 2011, \apj, 731, 54

\bibitem[{{Pawlik} {et~al.}(2013){Pawlik}, {Milosavljevi{\'c}}, \&
  {Bromm}}]{Pawlik13}
---. 2013, \apj, 767, 59

\bibitem[{{Planck Collaboration} {et~al.}(2016){Planck Collaboration}, {Ade},
  {Aghanim}, {Arnaud}, {Ashdown}, {Aumont}, {Baccigalupi}, {Banday},
  {Barreiro}, {Bartlett}, \& et~al.}]{Planck16}
{Planck Collaboration}, {Ade}, P.~A.~R., {Aghanim}, N., {Arnaud}, M.,
  {Ashdown}, M., {Aumont}, J., {Baccigalupi}, C., {Banday}, A.~J., {Barreiro},
  R.~B., {Bartlett}, J.~G., \& et~al. 2016, \aap, 594, A13

\bibitem[{{Pontzen} \& {Governato}(2012)}]{Pontzen12}
{Pontzen}, A., \& {Governato}, F. 2012, \mnras, 421, 3464

\bibitem[{{Ricotti} {et~al.}(2016){Ricotti}, {Parry}, \& {Gnedin}}]{Ricotti16}
{Ricotti}, M., {Parry}, O.~H., \& {Gnedin}, N.~Y. 2016, \apj, 831, 204

\bibitem[{{Robertson} {et~al.}(2015){Robertson}, {Ellis}, {Furlanetto}, \&
  {Dunlop}}]{Robertson15}
{Robertson}, B.~E., {Ellis}, R.~S., {Furlanetto}, S.~R., \& {Dunlop}, J.~S.
  2015, \apjl, 802, L19

\bibitem[{{Robertson} \& {Kravtsov}(2008)}]{Robertson08}
{Robertson}, B.~E., \& {Kravtsov}, A.~V. 2008, \apj, 680, 1083

\bibitem[{{Romano-D{\'{\i}}az} {et~al.}(2014){Romano-D{\'{\i}}az}, {Shlosman},
  {Choi}, \& {Sadoun}}]{Romano-Diaz14}
{Romano-D{\'{\i}}az}, E., {Shlosman}, I., {Choi}, J.-H., \& {Sadoun}, R. 2014,
  \apjl, 790, L32

\bibitem[{{Scannapieco} {et~al.}(2012){Scannapieco}, {Wadepuhl}, {Parry},
  {Navarro}, {Jenkins}, {Springel}, {Teyssier}, {Carlson}, {Couchman}, {Crain},
  {Dalla Vecchia}, {Frenk}, {Kobayashi}, {Monaco}, {Murante}, {Okamoto},
  {Quinn}, {Schaye}, {Stinson}, {Theuns}, {Wadsley}, {White}, \&
  {Woods}}]{Scannapieco12}
{Scannapieco}, C., {Wadepuhl}, M., {Parry}, O.~H., {Navarro}, J.~F., {Jenkins},
  A., {Springel}, V., {Teyssier}, R., {Carlson}, E., {Couchman}, H.~M.~P.,
  {Crain}, R.~A., {Dalla Vecchia}, C., {Frenk}, C.~S., {Kobayashi}, C.,
  {Monaco}, P., {Murante}, G., {Okamoto}, T., {Quinn}, T., {Schaye}, J.,
  {Stinson}, G.~S., {Theuns}, T., {Wadsley}, J., {White}, S.~D.~M., \& {Woods},
  R. 2012, \mnras, 423, 1726

\bibitem[{{Schaye} {et~al.}(2015){Schaye}, {Crain}, {Bower}, {Furlong},
  {Schaller}, {Theuns}, {Dalla Vecchia}, {Frenk}, {McCarthy}, {Helly},
  {Jenkins}, {Rosas-Guevara}, {White}, {Baes}, {Booth}, {Camps}, {Navarro},
  {Qu}, {Rahmati}, {Sawala}, {Thomas}, \& {Trayford}}]{Schaye15}
{Schaye}, J., {Crain}, R.~A., {Bower}, R.~G., {Furlong}, M., {Schaller}, M.,
  {Theuns}, T., {Dalla Vecchia}, C., {Frenk}, C.~S., {McCarthy}, I.~G.,
  {Helly}, J.~C., {Jenkins}, A., {Rosas-Guevara}, Y.~M., {White}, S.~D.~M.,
  {Baes}, M., {Booth}, C.~M., {Camps}, P., {Navarro}, J.~F., {Qu}, Y.,
  {Rahmati}, A., {Sawala}, T., {Thomas}, P.~A., \& {Trayford}, J. 2015, \mnras,
  446, 521

\bibitem[{{Schaye} \& {Dalla Vecchia}(2008)}]{Schaye08}
{Schaye}, J., \& {Dalla Vecchia}, C. 2008, \mnras, 383, 1210

\bibitem[{{Schaye} {et~al.}(2010){Schaye}, {Dalla Vecchia}, {Booth}, {Wiersma},
  {Theuns}, {Haas}, {Bertone}, {Duffy}, {McCarthy}, \& {van de
  Voort}}]{Schaye10}
{Schaye}, J., {Dalla Vecchia}, C., {Booth}, C.~M., {Wiersma}, R.~P.~C.,
  {Theuns}, T., {Haas}, M.~R., {Bertone}, S., {Duffy}, A.~R., {McCarthy},
  I.~G., \& {van de Voort}, F. 2010, \mnras, 402, 1536

\bibitem[{{Shibuya} {et~al.}(2012){Shibuya}, {Kashikawa}, {Ota}, {Iye},
  {Ouchi}, {Furusawa}, {Shimasaku}, \& {Hattori}}]{Shibuya12}
{Shibuya}, T., {Kashikawa}, N., {Ota}, K., {Iye}, M., {Ouchi}, M., {Furusawa},
  H., {Shimasaku}, K., \& {Hattori}, T. 2012, \apj, 752, 114

\bibitem[{{Springel}(2005)}]{Springel05e}
{Springel}, V. 2005, MNRAS, 364, 1105

\bibitem[{{Susa} \& {Umemura}(2004)}]{Susa04}
{Susa}, H., \& {Umemura}, M. 2004, ApJL, 610, L5

\bibitem[{{Tacconi} {et~al.}(2013){Tacconi}, {Neri}, {Genzel}, {Combes},
  {Bolatto}, {Cooper}, {Wuyts}, {Bournaud}, {Burkert}, {Comerford}, {Cox},
  {Davis}, {F{\"o}rster Schreiber}, {Garc{\'{\i}}a-Burillo}, {Gracia-Carpio},
  {Lutz}, {Naab}, {Newman}, {Omont}, {Saintonge}, {Shapiro Griffin}, {Shapley},
  {Sternberg}, \& {Weiner}}]{Tacconi13}
{Tacconi}, L.~J., {Neri}, R., {Genzel}, R., {Combes}, F., {Bolatto}, A.,
  {Cooper}, M.~C., {Wuyts}, S., {Bournaud}, F., {Burkert}, A., {Comerford}, J.,
  {Cox}, P., {Davis}, M., {F{\"o}rster Schreiber}, N.~M.,
  {Garc{\'{\i}}a-Burillo}, S., {Gracia-Carpio}, J., {Lutz}, D., {Naab}, T.,
  {Newman}, S., {Omont}, A., {Saintonge}, A., {Shapiro Griffin}, K., {Shapley},
  A., {Sternberg}, A., \& {Weiner}, B. 2013, \apj, 768, 74

\bibitem[{{Tanvir} {et~al.}(2012){Tanvir}, {Levan}, {Fruchter}, {Fynbo},
  {Hjorth}, {Wiersema}, {Bremer}, {Rhoads}, {Jakobsson}, {O'Brien}, {Stanway},
  {Bersier}, {Natarajan}, {Greiner}, {Watson}, {Castro-Tirado}, {Wijers},
  {Starling}, {Misra}, {Graham}, \& {Kouveliotou}}]{Tanvir12}
{Tanvir}, N.~R., {Levan}, A.~J., {Fruchter}, A.~S., {Fynbo}, J.~P.~U.,
  {Hjorth}, J., {Wiersema}, K., {Bremer}, M.~N., {Rhoads}, J., {Jakobsson}, P.,
  {O'Brien}, P.~T., {Stanway}, E.~R., {Bersier}, D., {Natarajan}, P.,
  {Greiner}, J., {Watson}, D., {Castro-Tirado}, A.~J., {Wijers}, R.~A.~M.~J.,
  {Starling}, R.~L.~C., {Misra}, K., {Graham}, J.~F., \& {Kouveliotou}, C.
  2012, \apj, 754, 46

\bibitem[{{Watson} {et~al.}(2015){Watson}, {Christensen}, {Knudsen}, {Richard},
  {Gallazzi}, \& {Micha{\l}owski}}]{Watson15}
{Watson}, D., {Christensen}, L., {Knudsen}, K.~K., {Richard}, J., {Gallazzi},
  A., \& {Micha{\l}owski}, M.~J. 2015, \nat, 519, 327

\bibitem[{{Wise} {et~al.}(2012{\natexlab{a}}){Wise}, {Abel}, {Turk}, {Norman},
  \& {Smith}}]{Wise12b}
{Wise}, J.~H., {Abel}, T., {Turk}, M.~J., {Norman}, M.~L., \& {Smith}, B.~D.
  2012{\natexlab{a}}, \mnras, 427, 311

\bibitem[{{Wise} {et~al.}(2014){Wise}, {Demchenko}, {Halicek}, {Norman},
  {Turk}, {Abel}, \& {Smith}}]{Wise14}
{Wise}, J.~H., {Demchenko}, V.~G., {Halicek}, M.~T., {Norman}, M.~L., {Turk},
  M.~J., {Abel}, T., \& {Smith}, B.~D. 2014, \mnras, 442, 2560

\bibitem[{{Wise} {et~al.}(2012{\natexlab{b}}){Wise}, {Turk}, {Norman}, \&
  {Abel}}]{Wise12a}
{Wise}, J.~H., {Turk}, M.~J., {Norman}, M.~L., \& {Abel}, T.
  2012{\natexlab{b}}, \apj, 745, 50

\bibitem[{{Yajima} {et~al.}(2011){Yajima}, {Choi}, \& {Nagamine}}]{Yajima11}
{Yajima}, H., {Choi}, J.-H., \& {Nagamine}, K. 2011, \mnras, 412, 411

\bibitem[{{Yajima} {et~al.}(2012){Yajima}, {Choi}, \& {Nagamine}}]{Yajima12h}
---. 2012, \mnras, 427, 2889

\bibitem[{{Yajima} {et~al.}(2015{\natexlab{a}}){Yajima}, {Li}, {Zhu}, \&
  {Abel}}]{Yajima15a}
{Yajima}, H., {Li}, Y., {Zhu}, Q., \& {Abel}, T. 2015{\natexlab{a}}, \apj, 801,
  52

\bibitem[{{Yajima} {et~al.}(2014{\natexlab{a}}){Yajima}, {Li}, {Zhu}, {Abel},
  {Gronwall}, \& {Ciardullo}}]{Yajima14c}
{Yajima}, H., {Li}, Y., {Zhu}, Q., {Abel}, T., {Gronwall}, C., \& {Ciardullo},
  R. 2014{\natexlab{a}}, \mnras, 440, 776

\bibitem[{{Yajima} {et~al.}(2014{\natexlab{b}}){Yajima}, {Nagamine},
  {Thompson}, \& {Choi}}]{Yajima14b}
{Yajima}, H., {Nagamine}, K., {Thompson}, R., \& {Choi}, J.-H.
  2014{\natexlab{b}}, \mnras, 439, 3073

\bibitem[{{Yajima} {et~al.}(2015{\natexlab{b}}){Yajima}, {Shlosman},
  {Romano-D{\'{\i}}az}, \& {Nagamine}}]{Yajima15c}
{Yajima}, H., {Shlosman}, I., {Romano-D{\'{\i}}az}, E., \& {Nagamine}, K.
  2015{\natexlab{b}}, \mnras, 451, 418

\bibitem[{{Yajima} {et~al.}(2017){Yajima}, {Sugimura}, \&
  {Hasegawa}}]{Yajima17}
{Yajima}, H., {Sugimura}, K., \& {Hasegawa}, K. 2017, ArXiv e-prints

\bibitem[{{Yajima} {et~al.}(2009){Yajima}, {Umemura}, {Mori}, \&
  {Nakamoto}}]{Yajima09}
{Yajima}, H., {Umemura}, M., {Mori}, M., \& {Nakamoto}, T. 2009, MNRAS, 398,
  715

\bibitem[{{Zhu} {et~al.}(2016){Zhu}, {Marinacci}, {Maji}, {Li}, {Springel}, \&
  {Hernquist}}]{Zhu16}
{Zhu}, Q., {Marinacci}, F., {Maji}, M., {Li}, Y., {Springel}, V., \&
  {Hernquist}, L. 2016, \mnras, 458, 1559

\bibitem[{{Zitrin} {et~al.}(2015){Zitrin}, {Labb{\'e}}, {Belli}, {Bouwens},
  {Ellis}, {Roberts-Borsani}, {Stark}, {Oesch}, \& {Smit}}]{Zitrin15}
{Zitrin}, A., {Labb{\'e}}, I., {Belli}, S., {Bouwens}, R., {Ellis}, R.~S.,
  {Roberts-Borsani}, G., {Stark}, D.~P., {Oesch}, P.~A., \& {Smit}, R. 2015,
  \apjl, 810, L12

\end{thebibliography}

\label{lastpage}

\end{document}